\documentclass[aps,prapplied,twocolumn,superscriptaddress]{revtex4-2}

\usepackage{graphicx}
\usepackage{subfigure}
\usepackage{amsmath}
\usepackage{amssymb}
\usepackage{bm}
\usepackage[colorlinks=true,citecolor=blue,urlcolor=blue,linkcolor=blue,hyperfigures=true]{hyperref}
\usepackage{comment}
\usepackage{dcolumn}

\def\Eq{Eq.~}
\def\Eqs{Eqs.~}
\def\Fig{Fig.~}
\def\Figs{Figs.~}
\def\Ref{Ref.~}

\def\App{Appendix~}
\def\be{\begin{equation}}
\def\ee{\end{equation}}
\def\bea{\begin{eqnarray}}
\def\eea{\end{eqnarray}}

\def\ie{\textit{i.e.}~}
\def\eg{\textit{e.g.}~}

\def\dd{\mathrm{d}}
\def\x10{\times 10}

\newcommand{\ket}[1]{\left| #1 \right\rangle}

\begin{document}

\title{Carrier-Suppressed Multiple Single-Sideband Laser Source\\for Atom Cooling and Interferometry}

\author{S. Templier}
\affiliation{iXblue, 34 rue de la Croix de Fer, 78105 Saint-Germain-en-Laye, France}
\affiliation{LP2N, Laboratoire Photonique, Num\'{e}rique et Nanosciences, Universit\'{e} Bordeaux--IOGS--CNRS:UMR 5298, 1 rue Fran\c{c}ois Mitterrand, 33400 Talence, France}

\author{J. Hauden}
\author{P. Cheiney}
\author{F. Napolitano}
\author{H. Porte}
\affiliation{iXblue, 34 rue de la Croix de Fer, 78105 Saint-Germain-en-Laye, France}

\author{P. Bouyer}
\affiliation{LP2N, Laboratoire Photonique, Num\'{e}rique et Nanosciences, Universit\'{e} Bordeaux--IOGS--CNRS:UMR 5298, 1 rue Fran\c{c}ois Mitterrand, 33400 Talence, France}

\author{B. Barrett}
\affiliation{iXblue, 34 rue de la Croix de Fer, 78105 Saint-Germain-en-Laye, France}
\affiliation{LP2N, Laboratoire Photonique, Num\'{e}rique et Nanosciences, Universit\'{e} Bordeaux--IOGS--CNRS:UMR 5298, 1 rue Fran\c{c}ois Mitterrand, 33400 Talence, France}
\affiliation{Department of Physics, University of New Brunswick, 8 Bailey Drive, PO Box 4400, Fredericton NB, E3B 5A3, Canada}

\author{B. Battelier}
\email[Corresponding author: ]{baptiste.battelier@institutoptique.fr}
\affiliation{LP2N, Laboratoire Photonique, Num\'{e}rique et Nanosciences, Universit\'{e} Bordeaux--IOGS--CNRS:UMR 5298, 1 rue Fran\c{c}ois Mitterrand, 33400 Talence, France}

\date{\today}

\begin{abstract}
We present a new electro-optic modulation technique that enables a single laser diode to realize a cold-atom source and a quantum inertial sensor based on matter-wave interferometry. Using carrier-suppressed dual single-sideband modulation, an IQ modulator generates two optical sidebands from separate radio-frequency (rf) signals. These sidebands are controlled independently in frequency, phase, and power using standard rf components. Our laser source exhibits improved rejection of parasitic sidebands compared to those based on phase modulators, which generate large systematic shifts in atom interferometers. We measure the influence of residual laser lines on an atom-interferometric gravimeter and show agreement with a theoretical model. We estimate a reduction of the systematic shift by two orders of magnitude compared to previous architectures, and reach a long-term sensitivity of 15 n$g$ on the gravitational acceleration with an interrogation time of only $T = 20$ ms. Finally, we characterize the performance of our integrated laser system, and show that it is suitable for mobile sensing applications including gravity surveys and inertial navigation.
\end{abstract}

\maketitle

\section{Introduction}
\label{sec:Introduction}

The advent of laser cooling and atom interferometry \cite{Berman1997, Cronin2009} has hailed new classes of inertial sensors with unprecedented sensitivity, including gravimeters \cite{Kasevich1991, LeGouet2008, Freier2016}, gravity gradiometers \cite{Sorrentino2014, Wang2017}, and gyroscopes \cite{Gustavson1997, Dutta2016}. However, these instruments are typically large, complex laboratory experiments that require quiet, stable conditions to operate reliably. Working in the field or onboard vehicles demands a drastic reduction in the size, weight and power consumption of these devices \cite{Rushton2014, Barrett2016, Bidel2018, Becker2018}, while also requiring a high level of robustness and resistance to environmental disturbances \cite{Zhang2018}. Over the past few years, significant progress has been made in the commercialization of cold-atom-based sensors \cite{Menoret2018}, and it is widely believed these high-performance inertial sensors will lead to a breakthrough in technology for many fields, including geophysics \cite{Canuel2018}, metrology \cite{Parker2018}, and inertial navigation \cite{Cheiney2018}. Laser sources represent one of the most crucial and complex parts of these systems. They require at least two frequencies (\eg for laser-cooling alkali metals or inducing Raman transitions), with an accuracy of $\sim 100$ kHz. A typical atom interferometer sequence also demands tunability over a range of $\sim 1$ GHz near the atomic transition, and sub-ms response times. Moreover, for a Raman interferometer, the two optical frequencies must be phase coherent, with a relative phase noise $\lesssim 60$ dBc/Hz at an offset frequency of 10 Hz.

Telecom frequency-based architectures afford the strength of extensive development with reliable and robust off-the-shelf components and can be frequency doubled using a periodically-poled lithium-niobate (PPLN) crystal \cite{Sabulsky2020}. Within the scope of all-fibered laser systems designed for laser cooling and atom interferometry, several configurations have been demonstrated. The use of two separate laser diodes, one operating at each frequency, requires an optical phase lock \cite{Oh2016, Li2017}. Frequency offset master-slave architectures offer real-time frequency agility by controlling the current of the slave diode \cite{Menoret2011}, but are typically limited in bandwidth and dynamic range by the electronics. Electro-optically generated sidebands in a servo-locked system can also achieve good frequency agility \cite{Peng2014, Theron2015, Battelier2016,Macrae2020}. To simplify such architectures, electro-optic phase modulation has been used to generate comb-like spectra controlled by an rf source \cite{Bonnin2013}. However, the use of pure phase modulation results in parasitic sidebands that are detrimental to atom interferometers, as they produce both a systematic measurement bias and spatial variations in fringe contrast \cite{Carraz2012, Wang2017}. Single-sideband (SSB) modulation using an IQ modulator has been shown to strongly reduce the effect of parasitic lines \cite{Zhu2018}. There, however, the tunability of the laser source was realized by controlling the carrier frequency of the laser diode.

In this work, we demonstrate a new 780 nm laser architecture based on carrier-suppressed dual single-sideband (CS-DSSB) modulation in an all-fibered IQ modulator operating at 1560 nm. Here, electro-optic modulation techniques are utilized to generate two ``principal'' optical sidebands, while suppressing the carrier frequency and other parasitic lines. These principal sidebands are controlled in frequency, phase and power in the rf domain---affording all the benefits of modern rf sources in terms of agility, stability and response time. With this system, we realize a complete interferometer sequence (including atom trap loading, sub-Doppler cooling, state preparation, interrogation, and detection) using only rf modulation while holding the carrier frequency of the laser fixed. Using this system, we realize a quantum gravimeter with a long-term stability $< 20$ n$g$ ($< 1$ mrad) using a modest interrogation time of $T = 20$ ms. Finally, we present a general model for the systematic bias of atomic gravimeters in the presence of residual laser lines. Our model is consistent with experimental measurements, and highlights for the first time that laser architectures with reduced parasitic laser lines (such as those based on based on IQ modulators \cite{Zhu2018}) can have non-negligible effects on the phase and contrast of atom interferometers.

\section{Laser architecture}
\label{sec:Architecture}

\begin{figure}[!tb]
  \centering
  \includegraphics[width=0.49\textwidth]{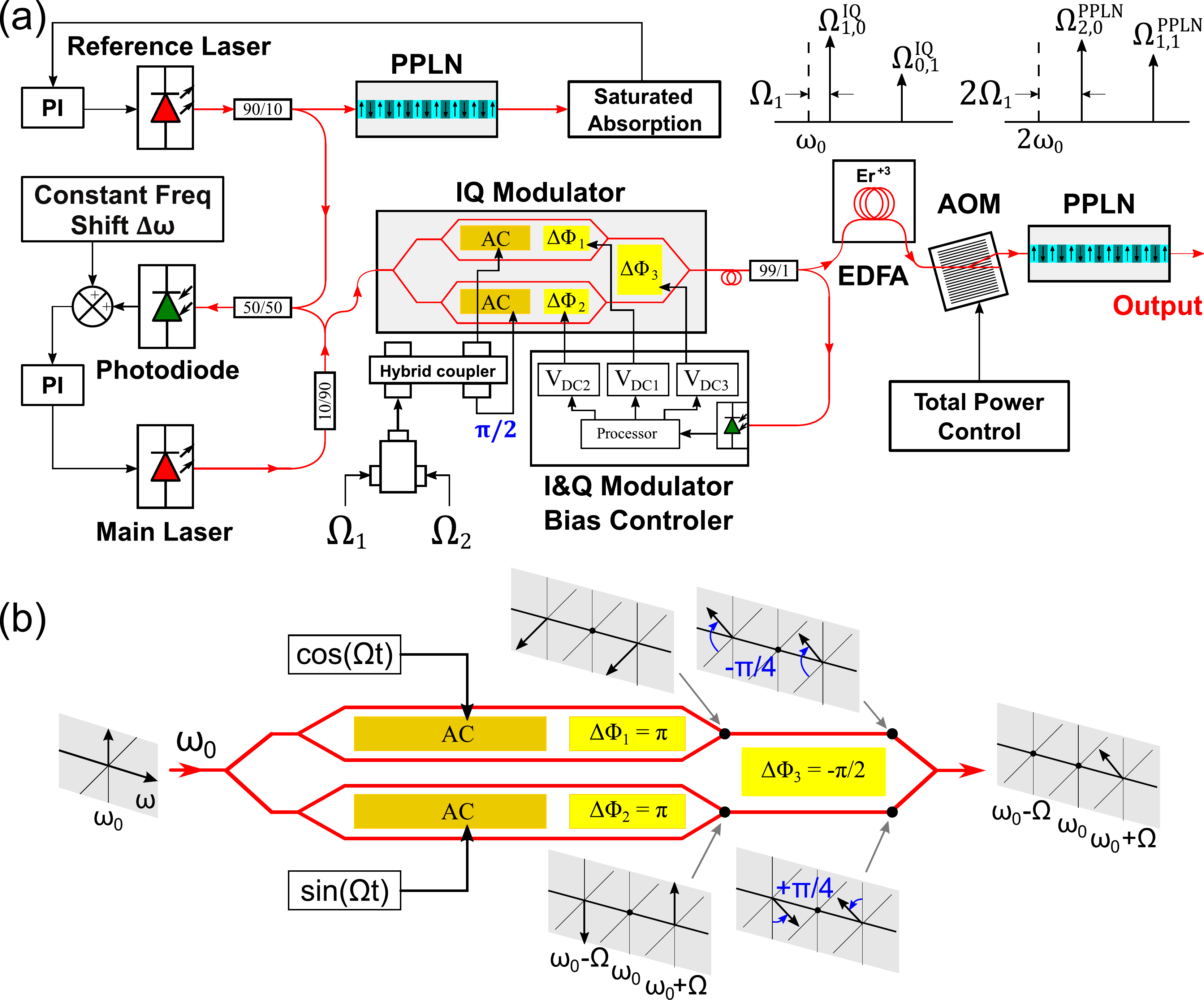}
  \caption{(a) Architecture of the laser source. PI: proportional-integrator lock. (b) Schematic of the electro-optic IQ modulator where, for clarity, only one frequency $\Omega$ is injected into the hybrid coupler. EDFA: erbium-doped fiber amplifier; AOM: acousto-optic modulator; PPLN: periodically-poled lithium niobate waveguide.}
  \label{fig:Architecture}
\end{figure}

The architecture of the laser system is presented in \Fig \ref{fig:Architecture}(a). A reference laser diode is locked to the $\ket{F = 2} \to \ket{F'=(2,3)}$ crossover transition of $^{87}$Rb using saturated absorption spectroscopy. The main laser (Redfern Integrated Optics external-cavity diode laser at 1560 nm, 20 mW output power) is frequency locked to the red of the reference laser via an optical beatnote at $\Delta\omega \simeq 2\pi \times 1.5$ GHz. For the laser-cooling phase, two principal sidebands are generated on the main laser at frequencies \(\Omega_1 = \Delta\omega + (\delta_{\rm CO} + \Delta_{\rm C})/2\) for cooling and $\Omega_2 = \Omega_1 + (\delta_{\rm HF} - \delta_{32})$ for repumping, both of which are blue-shifted from the reference laser. Here, $\delta_{32}/2\pi \simeq 266$ MHz is the splitting between excited states $\ket{F' = 2}$ and $\ket{F' = 3}$, $\delta_{\rm CO} = \delta_{32}/2$ is the detuning of the crossover resonance from $\ket{F' = 3}$, $\delta_{\rm HF}/2\pi = 6.834$ GHz is the ground-state hyperfine splitting, and $\Delta_{\rm C}/2\pi \simeq -20$ MHz is the detuning from $\ket{F' = 3}$ used for laser cooling. Similarly, for Raman interferometry, $\Omega_1 = \Delta\omega +(\Delta_{\rm R}-\delta_{\rm CO})/2$ and $\Omega_2 = \Omega_1 + \delta_{\rm HF}$, where $\Delta_{\rm R}$ is the detuning of the Raman beams from $\ket{F = 2} \to \ket{F' = 2}$. When in CS-DSSB mode, the optical loss in the IQ modulator ($\sim 12$ dB in our case) depends on the total injected rf power. The remaining 500 $\mu$W of optical power is sufficient to saturate the double-stage erbium-doped fiber amplifier (EDFA, Lumibird CEFA-C-PB-HP) after the modulator. The EDFA then outputs 2 W at 1560 nm with a $\sim 1\%$ power stability. This light subsequently undergoes second-harmonic generation (SHG) to 780 nm in a PPLN crystal waveguide (NTT Electronics WH-0780-000-F-B-C). A fibered acousto-optic modulator (AOM, Gooch \& Housego A35080-S-1/50-p4k7u), located between the EDFA and the PPLN waveguide, controls the total output power, which is 485 mW at maximum.  The optical power stability is approximately 0.5\% over $10^{4}$ s [see \Fig \ref{fig:Stability}(b)].

The telecom-domain IQ modulator (iXblue MXIQER-LN-30) consists of three optically-guided Mach-Zehnder interferometers (MZIs): two sub-MZIs nested inside a main one, as shown in \Fig \ref{fig:Architecture}(b). Each MZI has a broad modulation bandwidth of 30 GHz. An rf signal containing the two frequencies $\Omega_1$ and $\Omega_2$ is sent through a hybrid coupler, which equally splits the signal and phase shifts one arm by $\pi/2$. The resulting two rf signals are sent to the AC electrodes of each sub-MZI. A commercial bias voltage controller (iXblue IQ-MBC-LAB) delivers three continuous voltages to the DC electrodes [shown in yellow in \Fig \ref{fig:Architecture}(b)] to control and stabilize in real time the bias phases of each MZI ($\Delta\Phi_1$, $\Delta\Phi_2$, and $\Delta\Phi_3$). This DC bias lock is realized by modulating the light passing through each sub-MZI at 1 kHz, and deriving an error signal from the FFT of the corresponding optical signals. This control is critical since these phases can drift dramatically due to temperature sensitivity and internal charge dynamics.

The primary function of each sub-MZI is to suppress the optical carrier frequency. We emphasize that the global phase shifts $\Delta\Phi_1$ and $\Delta\Phi_2 = \pm\pi$ guarantees destructive interference of the carrier, while the input rf phase $\varphi$ determines the relative phase between the two remaining sidebands. A schematic of the IQ modulator, operating in carrier-suppressed single-sideband (CS-SSB) mode \cite{Izutsu1981, Shimotsu2001}, is shown in \Fig \ref{fig:Architecture}(b). Here, for clarity, only one frequency $\Omega$ is injected into the hybrid coupler, which then modulates the two sub-MZIs with $\cos(\Omega t)$ and $\sin(\Omega t)$, respectively. This results in carrier-suppressed optical signals which are in-quadrature with one another. These two signals are then combined in the main MZI, where the bias phase $\Delta\Phi_3$ determines the surviving harmonics. Specifically, the upper sideband (order $+1$) remains when $\Delta\Phi_3 = -\pi/2$, whereas the lower sideband (order $-1$) survives for $+\pi/2$. CS-DSSB modulation is completely analogous to CS-SSB modulation, except two separate rf signals are injected into the IQ modulator---generating two independent sidebands. The relative phase between these optical signals is directly controlled by the rf source.

Both rf signals $\Omega_1$ and $\Omega_2$ are controlled at the sub-Hz level with a custom-built rf source. A voltage-controlled oscillator (VCO, Minicircuits ZX95-1750W-S+) is used to generate $\Omega_1$. A low-noise phase-locked dielectric resonator oscillator (PLDRO, Polaris SPLDRO-RE100-6800-P13-CP) generates $6.8$ GHz and is mixed with the VCO to produce $\Omega_2$. Similar to the architecture described in \Ref \citenum{Lautier2014}, the frequency difference $\Omega_2 - \Omega_1$ (and its associated phase) is controlled by a direct digital synthesizer (Analog Devices AD9959) mixed with the PLDRO. This scheme guarantees common-mode suppression of the VCO's phase noise. The power ratio between $\Omega_1$ and $\Omega_2$ is controlled using a variable-voltage rf attenuator---enabling us to suppress constant light shifts during the interferometer \cite{LeGouet2008}, and to optimize the optical molasses, state preparation, and detection phases of the atom interferometer sequence.

\begin{figure}[!tb]
  \centering
  \includegraphics[width=0.49\textwidth]{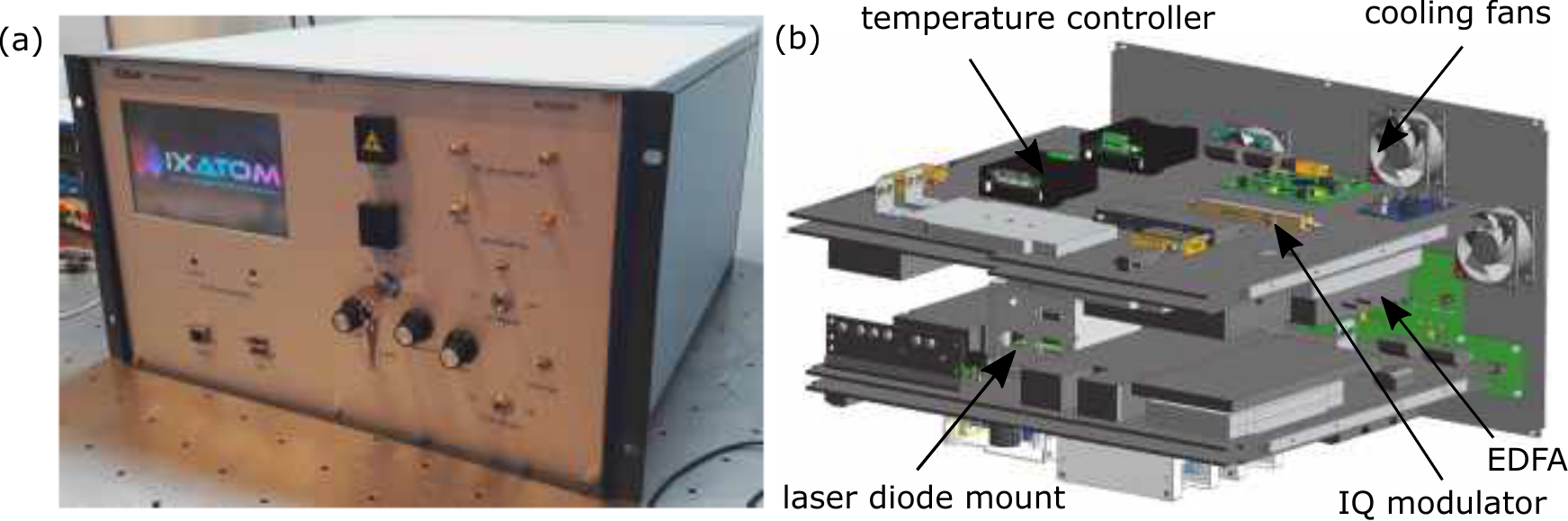}
  \caption{(a) Integrated laser source in a 19-inch 9U rack. (b) 3D design of the laser system, including cooling fans and active temperature controllers for individual components.}
  \label{fig:Modbox}
\end{figure}

The laser system is integrated into a 19-inch 9U rack, including a CPU and a graphical user interface, as shown in \Fig \ref{fig:Modbox}. A rubidium vapor cell is surrounded by heat-tape to maintain a homogeneous temperature ($\simeq 30^{\circ} $C). The temperature of the laser diodes and PPLN waveguides is actively controlled in independent containers for use in thermally-variable environments.

\section{Characterization of the laser system}
\label{sec:IQModulator}

\begin{figure}[!tb]
  \centering
  \includegraphics[width=0.49\textwidth]{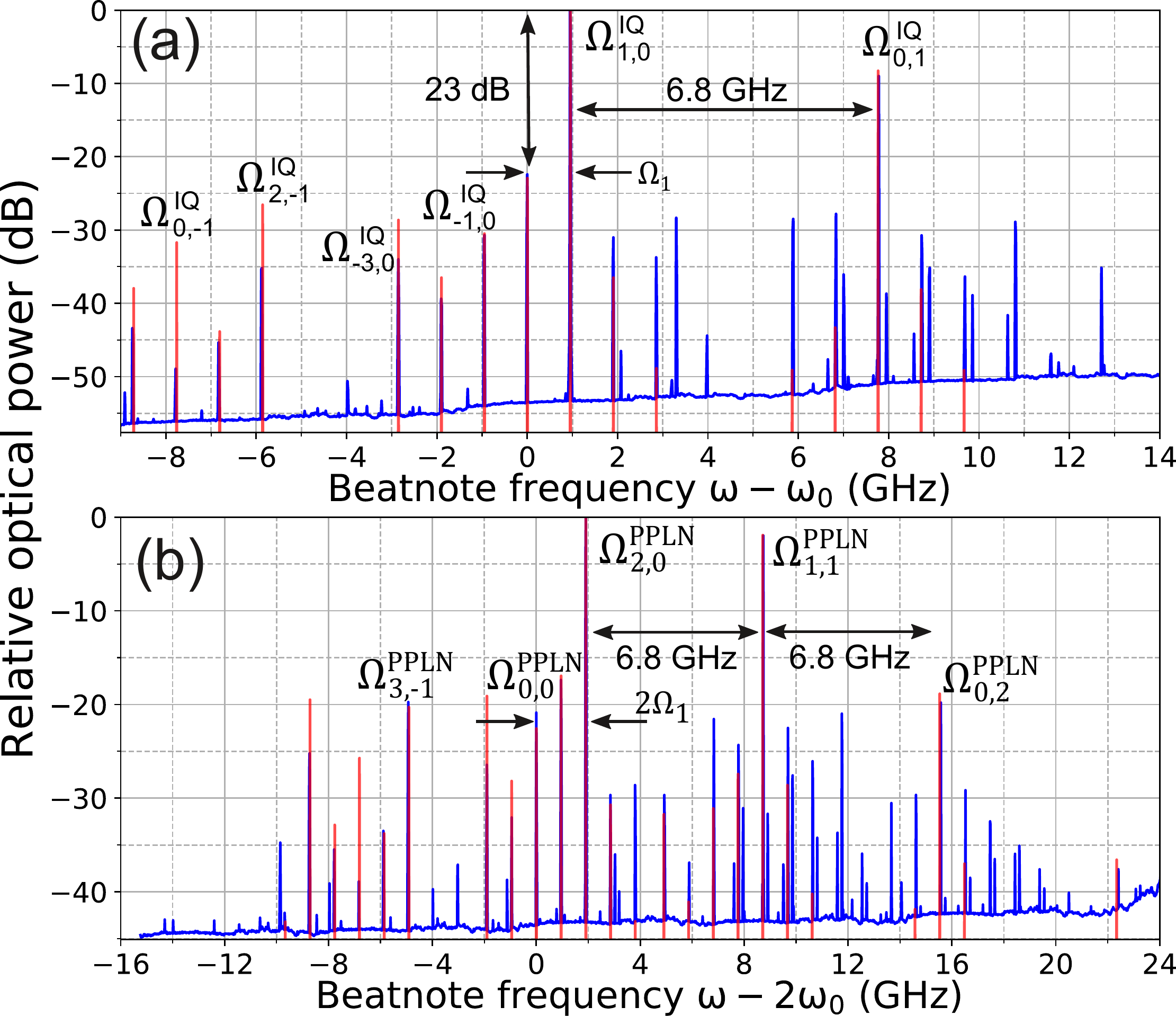}
  \caption{Frequency spectrum after CS-DSSB modulation at 1560 nm (a) and 780 nm (b). Blue lines show measured spectra during Raman interferometry obtained by beating with a stable local oscillator at 1560 nm and 780 nm, respectively. Red lines correspond to estimates from a numerical model, which are denoted $\Omega_{n,m}^{\rm IQ} = \omega_0 + n\Omega_1 + m\Omega_2$, where $n$ and $m$ are integers. Here, the principal sidebands are labelled $\Omega_{1,0}^{\rm IQ}$ and $\Omega_{0,1}^{\rm IQ}$. Similarly, at 780 nm, lines are labelled $\Omega_{N,M}^{\rm PPLN} = 2\omega_0 + N\Omega_1 + M\Omega_2$, where the principal sidebands are $\Omega_{2,0}^{\rm PPLN}$ and $\Omega_{1,1}^{\rm PPLN}$. The input rf power was 17 dBm and 9 dBm for $\Omega_1$ and $\Omega_2$, respectively, corresponding to modulation depths of 0.55 and 0.23.}
  \label{fig:IQSpectra}
\end{figure}

Figure \ref{fig:IQSpectra}(a) shows an example of the spectrum obtained after CS-DSSB modulation at 1560 nm using a beatnote with the reference laser \footnote{Due to the limited bandwidth of the photodiode ($\sim 10$ GHz), to obtain the spectra shown in \Fig \ref{fig:IQSpectra} we concatenated several frequency scans. In each scan, we shifted the local oscillator such that the peaks lied within the sensitivity band, and we renormalized the spectrum by the detector's transfer function. This had the effect of increasing the noise floor at higher frequencies.}. Here, we demonstrate suppression of the carrier $\omega_0$ by $\sim 23$ dB and all other parasitic lines below 25 dB. The two principal sidebands, labelled $\Omega_{1,0}^{\rm IQ} = \omega_0 + \Omega_1$ and $\Omega_{0,1}^{\rm IQ} = \omega_0 + \Omega_2$, are offset from the carrier by $\Omega_1/2\pi = 1$ GHz and are separated by $\Omega_2 - \Omega_1 = \delta_{\rm HF}$. Similarly, the optical spectrum at 780 nm (\ie after the PPLN) is shown in \Fig \ref{fig:IQSpectra}(b). Here, the principal laser lines used for cooling and interferometry are labelled $\Omega_{2,0}^{\rm PPLN} = 2\omega_0 + 2\Omega_1$ and $\Omega_{1,1}^{\rm PPLN} = 2\omega_0 + \Omega_1 + \Omega_2$. These two frequencies are generated by doubling $\Omega_{1,0}^{\rm IQ}$ and by summing $\Omega_{1,0}^{\rm IQ} + \Omega_{0,1}^{\rm IQ}$, respectively.

We have developed both an analytical model for the electric field (see \App \ref{sec:IQModel}) and a numerical model for the spectra, which is overlayed with the measurements shown in \Fig \ref{fig:IQSpectra}. This numerical model accurately reproduces several effects, such as the level of carrier suppression, which is determined by the accuracy of bias phases $\Delta\Phi_1$ and $\Delta\Phi_2$. The suppression of all other sidebands is influenced by a combination of the bias phase $\Delta\Phi_3$, and the phase and amplitude imbalance caused by the hybrid coupler. These imbalances are measured and input into the model. The IQ modulator can also exhibit several manufacturing defects, such as asymmetries between the arms of the MZIs and irregularities in the electrodes, which can produce additional lines. The PPLN then mixes all the lines through the process of SHG. Finally, we observe several lines in the spectra that are not predicted by the numerical model. These are due to spurious frequencies generated by the rf source that arise due to harmonic distortion, and mixing residues that are not completely suppressed by the rf filters within the source.

\begin{figure}[!tb]
  \centering
  \includegraphics[width=0.49\textwidth]{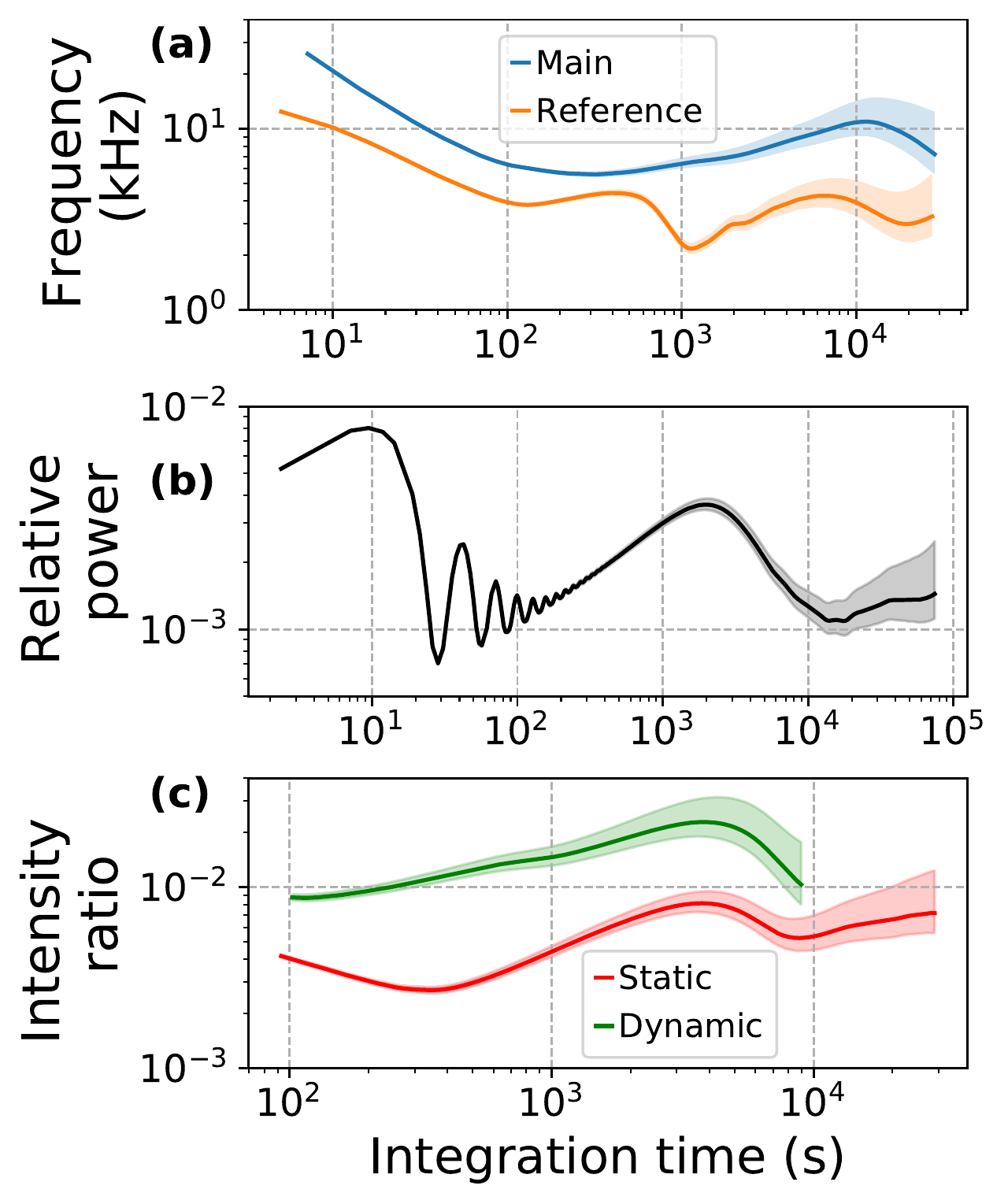}
  \caption{(a) Allan deviation of the optical beatnote between the reference laser (blue curve) and the IQ laser source (orange curve) and a third ``master'' laser at 1560 nm. (b) Allan deviation of the relative optical power emitted by the laser source. (c) Allan deviation of the intensity ratio between principal lines $\Omega_{1,1}^{\rm PPLN}$ and $\Omega_{2,0}^{\rm PPLN}$ in both static (red curve) and dynamic (green curve) modes. The amplitude of each line was measured on a spectrum analyzer by beating the laser output with a local oscillator. In the static mode the rf signals remained at fixed frequency and power, while the dynamic mode corresponds to a typical an AI sequence where the rf signals are modulated in amplitude and frequency. The shaded areas in all plots indicates $1\sigma$ uncertainty.}
  \label{fig:Stability}
\end{figure}

Figure \ref{fig:Stability} summarizes the stability of the laser system. Here, we monitored the beatnote of the main laser and the reference laser with a third ``master'' laser operating at 1560 nm \footnote{The stability of the master laser was separately determined to be $\sim 1.5$ kHz after $10^3$ s using a frequency comb.}. The frequency stability of the laser source is below 10 kHz after $2\times 10^4$ s at 1560 nm---limited only by temperature variations in the VCOs that produced a frequency drift of $\sim 15$ kHz. The relative power stability (measured after a fiber-splitting bench) is at the level of 0.2\% after operating for 52 hours. Finally, we tracked the intensity of the two principal lines $\Omega_{1,1}^{\rm PPLN}$ and $\Omega_{2,0}^{\rm PPLN}$ in static and dynamic operating modes. The stability of these lines contribute to time-varying light shifts in the atom interferometer, and give an indication of the stability of other lines. The intensity ratio between these lines reaches a stability below 1\% in static mode, and $\sim 2\%$ in dynamic mode after 2.5 hours. At this level, we estimate a phase stability of $\sim 0.75$ mrad (12 ng) for the atom interferometer (see \App \ref{sec:ResidualLineEffects}).

\section{Laser cooling and atom interferometry}
\label{sec:Interferometer}

We now discuss the results of laser-cooling and atom interferometry experiments carried out with this laser system. Our experimental setup was previously described in \Ref \citenum{Cheiney2018}. A vapor-loaded 3D magneto-optical trap accumulates $\sim 5 \times 10^8$ atoms in 250 ms. This is followed by a 10 ms gray-molasses stage using the D2 transition \cite{Rosi2018}, where the atoms are cooled to 2.5 $\mu$K. In our retro-reflected three-beam MOT, this technique is particularly effective compared to a standard red molasses because the beams are further detuned from the cycling transition---resulting in improved intensity balance between beams and lower cloud temperatures. During the gray molasses, cold atoms are coherently transferred to the dark state $\ket{F = 1}$. Post-molasses, a magnetic bias field of $\sim 70$ mG is turned on and the atoms are prepared in the magnetically-insensitive $\ket{F = 1, m_F = 0}$ state using two 100 $\mu$s pulses of co-propagating Raman light \cite{Templier2021}. During $\sim 300$ $\mu$s, the Raman frequency is swept across the two-photon Zeeman resonances---coherently transferring atoms from $\ket{F = 1, m_F = \pm 1}$ to $\ket{F = 2, m_F = \pm 1}$ where they are subsequently removed with a blow-away pulse near $\ket{F = 2} \to \ket{F' = 3}$. This preparation sequence is repeated three times, which enables us to reach 95\% purity in $\ket{F=1,m_F=0}$ with no residual heating. During each of these stages, the principal lines produced by the laser source are modulated in amplitude and shifted in frequency using only rf control.

The atom interferometer is configured as a three-pulse Mach-Zehnder-type gravimeter \cite{Kasevich1991, LeGouet2008, Freier2016, Menoret2018}, with vertically-oriented Raman beams that are retro-reflected by a mirror. In this configuration, the phase shift of the interferometer is $\Delta\Phi = (\alpha - k_{\rm eff} g) T^2$, where $\alpha$ is the chirp rate applied to one of the Raman sidebands, $k_{\rm eff} \simeq 4\pi/\lambda$ is the effective wavevector of Raman light at wavelength $\lambda = 780$ nm, $g$ is the gravitational acceleration, $T$ is the free-fall time between Raman pulses. For atom interferometry experiments, the two pairs of counter-propagating Raman beams have orthogonal linear polarizations to suppress velocity-insensitive co-propagating transitions. Using microwave components, we adjust the intensity ratio between the two Raman lines to minimize the AC Stark shift between the two ground states. We typically measure shifts $< 20$ Hz/(mW/cm$^2$). A magnetically-shielded mechanical accelerometer (Thales EMA 1000) monitors vibrations of the reference mirror, and is used to correct for vibration-induced phase noise \cite{LeGouet2008, Barrett2016, Templier2021}.

\begin{figure}[!tb]
  \centering
  \includegraphics[width=0.49\textwidth]{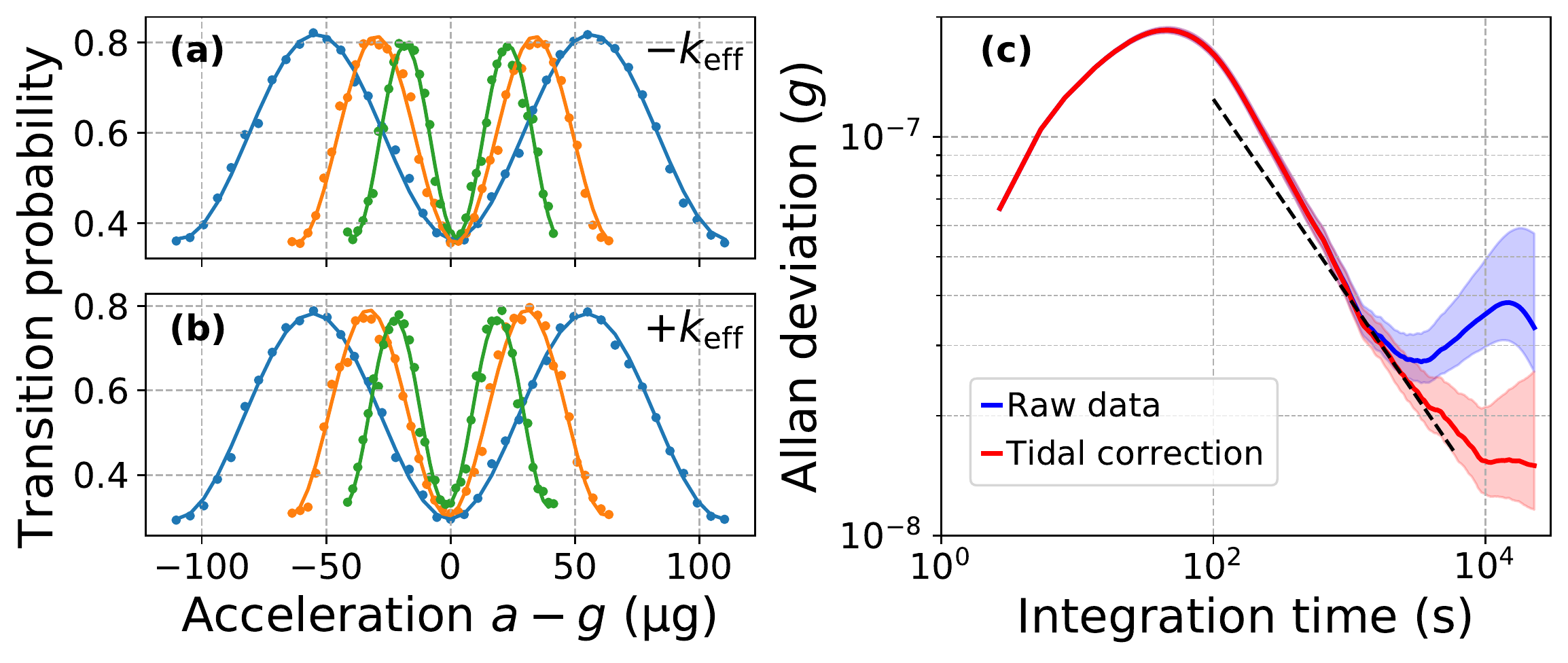}
  \caption{Interference fringes for opposite directions of momentum transfer (a) $-k_{\rm eff}$ and (b) $+k_{\rm eff}$. Data are shown for interrogation times $T = 19$ ms (blue), 25 ms (orange), and 31 ms (green). Solid lines are least-squares fits to the data. The horizontal axis is centered on the local reference of gravity $g = 9.805642$ m/s$^2$ \cite{Menoret2018}. (c) Allan deviation of raw acceleration measurements for $T = 20$ ms (blue curve), and after subtraction of tides (red curve). Shaded regions correspond to the $1\sigma$ statistical uncertainty in the Allan deviation, and the black dotted line is a fit of the form $\sigma_a/\sqrt{t}$. Other parameters: $\pi$-pulse duration $2\tau = 6$ $\mu$s; detuning $\Delta_{\rm R}/2\pi = -0.88$ GHz.}
  \label{fig:FringeADev}
\end{figure}

Figure \ref{fig:FringeADev} shows interference fringes obtained for different interrogation times. Here, the relative population in $\ket{F = 2}$ is measured as a function of $\alpha$, and the dark fringe common to all values of $T$ gives a measure of local gravity through $\alpha = k_{\rm eff} g$. To suppress non-inertial shifts due to the AC Stark and quadratic Zeeman effects, we combine measurements of $g$ with opposite wavevectors $\pm k_{\rm eff}$ \cite{LeGouet2008}. At $T = 31$ ms the fringes exhibit a contrast of $C = 0.43(1)$, offset noise $\sigma_P = 0.0079(1)$, contrast noise $\sigma_C = 0.0033(1)$, and phase noise of $\sigma_\phi = 0.141(4)$ rad/shot (limited by the self-noise of the mechanical accelerometer $\sim 360$ n$g/\sqrt{\rm Hz}$ \cite{Barrett2015}). These noise parameters are estimated by minimizing the negative log-likelihood distribution associated with the sinusoidal fit function---similar to the method presented in \Ref \citenum{Cheiney2018}. For a cycling time of $T_{\rm cyc} = 1.6$ s, we estimate a short-term acceleration sensitivity of $\sigma_a = \sigma_\phi \sqrt{T_{\rm cyc}}/k_{\rm eff} T^2 = 1.2$ $\mu g/\sqrt{\rm Hz}$.

To characterize the stability and ultimate sensitivity limit of our setup, we implemented a central fringe lock using a protocol similar that used in atomic clocks. Here, the phase of the interferometer is modulated between opposite mid-fringe positions ($\pm \pi/2$), where the sensitivity to phase fluctuations is largest \cite{Louchet-Chauvet2011}. We also alternated between momentum transfer directions ($\pm k_{\rm eff}$) and combined the measurements such that non-inertial effects are rejected on the fly. Hence, one lock cycle is complete every $T_{\rm lock} = 4 T_{\rm cyc} \simeq 6.4$ s. The chirp rate is steered onto the central fringe using a simple proportional-integrator feedback loop, with a time constant of $\sim 10 T_{\rm lock} = 64$ s. This scheme has the advantage of being insensitive to fringe offset and contrast variations, and optimizes the short-term sensitivity of the sensor. Figure \ref{fig:FringeADev}(c) shows the Allan deviation of acceleration measurements after tracking the central fringe for 15 hours. After a few time constants, the measurement sensitivity integrates as roughly $1/\sqrt{t}$, as expected for white Gaussian phase noise. From a fit to these data, we obtain a short-term sensitivity of $\sigma_a = 2.4$ $\mu g/\sqrt{\rm Hz}$---consistent with our estimates of the phase noise from the interference fringes. The tidal gravitational anomaly appears after $\sim 1$ hour, and is subtracted using a global model for tidal effects \cite{Timmen1995}. We reach a statistical precision of 15 n$g$ after $10^4$ s of integration---corresponding to a phase stability of $\sim 0.95$ mrad.

\begin{figure}[!tb]
  \centering
  \includegraphics[width=0.49\textwidth]{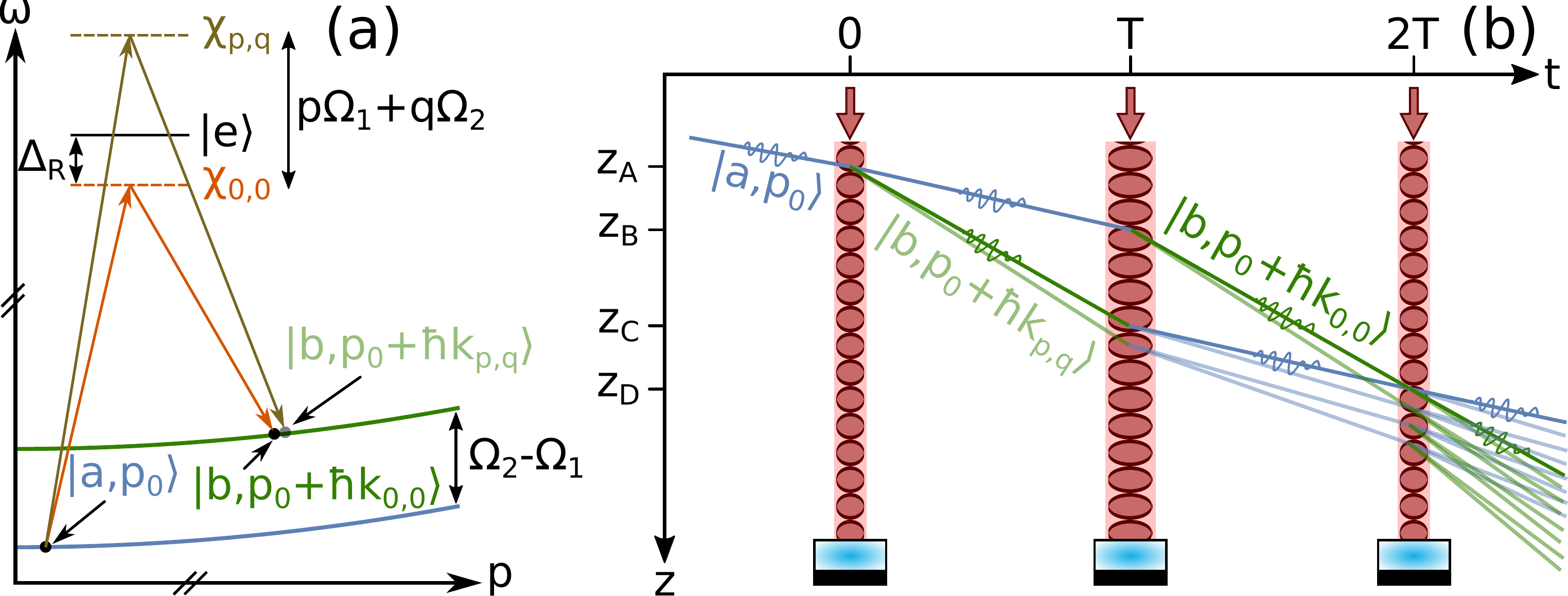}
  \caption{The effect of an additional laser line on an atom interferometer. (a) Two-photon Raman transitions between ground states $\ket{a}$ and $\ket{b}$ coupled by two pairs of laser lines; one with principal Rabi frequency $\chi_{0,0}$, and one residual $\chi_{p,q}$. The momentum transferred in each case is $\hbar k_{p,q} = \hbar(k_{\rm eff} + q\Delta k_1 + p\Delta k_2)$, where $\Delta k_1 = 2\Omega_1/c$, $\Delta k_2 = 2\Omega_2/c$, and $\Omega_2 - \Omega_1 = \delta_{\rm HF}$ is the splitting between ground states. (b) Illustration of the multi-path interferometer that results from an additional laser line. For sufficiently cold atoms, these pathways are within the de Broglie wavelength of the diffracted wavepackets---causing spatial interference that leads to a phase shift.}
  \label{fig:ParasiticAI}
\end{figure}

\section{Influence of residual lines}
\label{sec:ResidualLines}

The impact of additional laser lines on atom-interferometric sensors has previously been studied in the case of phase-modulated light in free space \cite{Carraz2012, Wang2017}, and more recently in an optical cavity \cite{Kristensen2021}. The presence of these lines adds several spatial components to the effective Rabi frequency for counter-propagating Raman transitions, resulting in a small spread of momenta transferred to the atom on each Raman pulse. This leads to a multi-path interferometer, as illustrated in \Fig \ref{fig:ParasiticAI}. Interference between the different momentum components leads to a spatially-varying Rabi frequency and phase shift, which depend on the relative line intensities and the position of the atoms relative to the reference mirror. The largest contributors to these effects are the two pairs of lines nearest the principal Raman pair. In \Fig \ref{fig:IQSpectra}(b), these lines correspond to $(\Omega_{3,-1}^{\rm PPLN}, \Omega_{2,0}^{\rm PPLN})$ and $(\Omega_{1,1}^{\rm PPLN}, \Omega_{0,2}^{\rm PPLN})$. In a phase modulator, the intensity of these lines is typically the same order of magnitude as the principal pair. With our method, we are able to suppress these lines at the 20 dB level, which strongly reduces their effects on the atom interferometer. Additionally, we show that the two Raman coupling associated with these residual lines have opposite signs, hence their contributions to the interferometer phase tend to cancel out (see \App \ref{sec:ResidualLineEffects}).

\begin{figure}[!tb]
  \centering
  \includegraphics[width=0.49\textwidth]{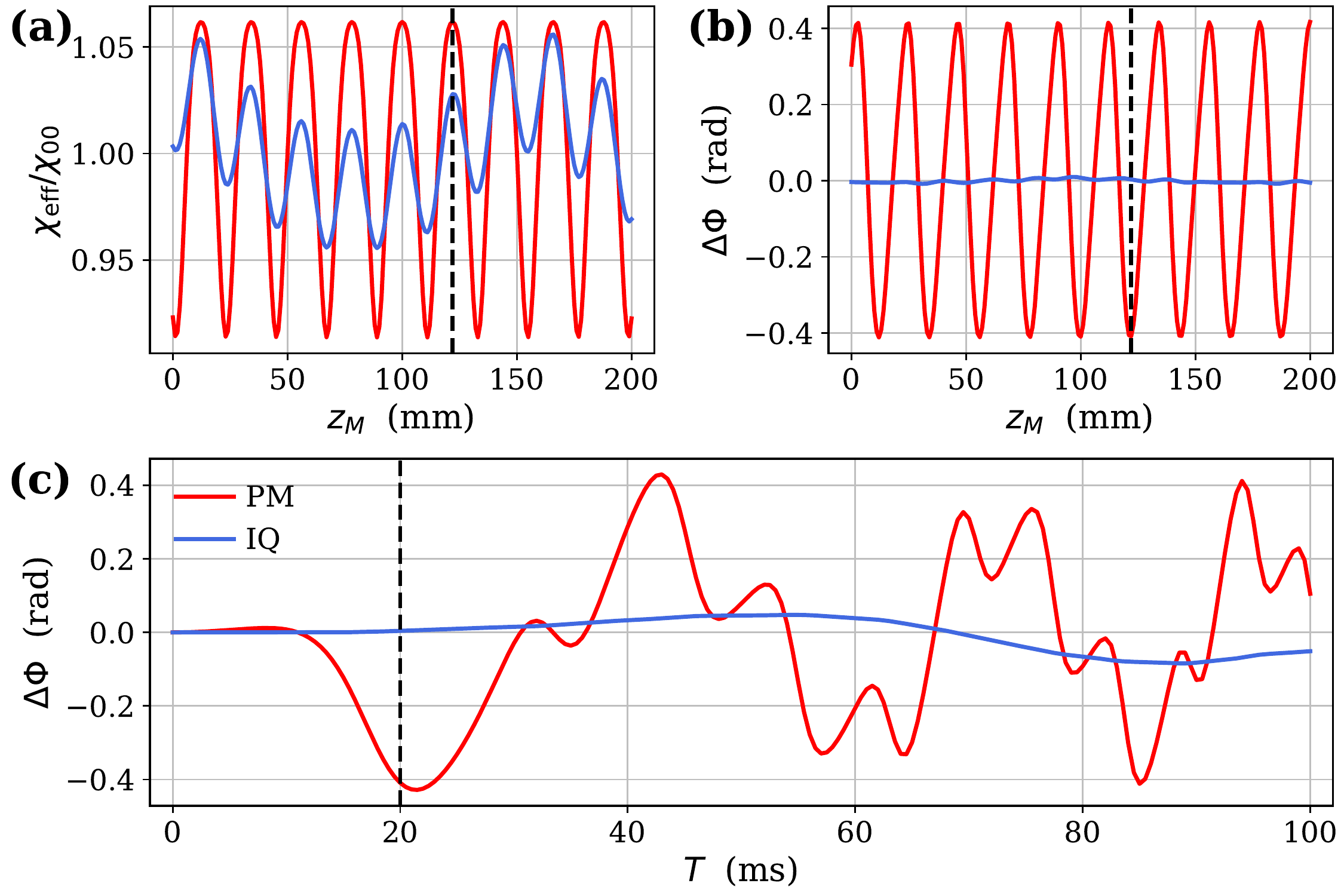}
  \caption{Comparison of the normalized Rabi frequency $\chi_{\rm eff}/\chi_{0,0}$ (a) and AI phase shift $\Delta\Phi$ (b,c) produced by a phase modulator (red curves) and an IQ modulator (blue curves). The phase shift is shown as a function of $z_{\rm M}$ (b) for fixed $T = 20$ ms, and as a function $T$ (c) for a fixed atom-mirror distance $z_{\rm M} = 122$ mm. IQ modulator line intensities are taken from \Fig \ref{fig:IQSpectra}(b). Phase modulator relative line intensities: $I_{0,m}/I_{0,0} = (-30, -12, -2, 0, -2, -12, -30)$ dB for modulation frequencies $m \delta_{\rm HF}$ with $m = -3,-2,\ldots,3$. Other AI parameters: TOF = 15 ms, $2\tau = 6$ $\mu$s, $\Delta_{\rm R} = -0.88$ GHz, initial velocity $v_0 = -15$ mm/s.}
  \label{fig:IQvsPM}
\end{figure}

Figure \ref{fig:IQvsPM} shows the predicted Rabi frequency and AI phase shift due to additional laser lines for the case of a standard phase modulator (PM) and an IQ modulator operating in CS-DSSB mode. The model for these effects is based on the multi-chromatic field produced by our laser system (see \App \ref{sec:IQModel}), along with measurements of the residual line intensities [\Fig \ref{fig:IQSpectra}(b)], and a precise determination of the initial atom-mirror distance $z_{\rm M}$ (see \App \ref{sec:MirrorDistance}). For the phase shift model, we also take into account the $k$-reversal process used in the experiment, which removes direction-independent systematic effects. Specifically, $\Delta\Phi = \frac{1}{2}(\Delta\phi_{\uparrow} - \Delta\phi_{\downarrow})$, where $\Delta\phi_{\uparrow\downarrow}$ is the phase shift for upward (downward) momentum kicks. From \Figs \ref{fig:IQvsPM}(a) and (b), both $\chi_{\rm eff}$ and $\Delta\Phi$ are periodic with the mirror position $z_{\rm M}$. The shape of these oscillations features a complex dependence on the relative lines intensities and their phase relationships. The key feature of the IQ modulator is that these modulation effects are highly suppressed compared to the phase modulator---especially for the spatial component corresponding to $\delta_{\rm HF}/2\pi = 6.834$ GHz (\ie spatial period of $21.9$ mm). An additional spatial modulation with period 157 mm is present in the IQ modulator due to residual lines at an offset frequency of $\Omega_1/2\pi = 0.955$ GHz. The phase shift due to the IQ modulator is primarily due to these lines. Figure \ref{fig:IQvsPM}(c) shows the phase shift as a function of the interrogation time $T$. Here, the IQ modulator offers several advantages in terms of systematic effects over a phase modulator. First, the amplitude of phase oscillations is reduced by more than a factor 5 up to $T = 100$ ms. Second, due to the larger spatial modulation period, $\Delta\Phi$ accumulates on a much longer scale and with a moderate variation. This reduces the level at which experimental parameters (\eg line intensities, Rabi frequency, and mirror position) must be known in order to accurately determine this systematic shift, or to operate where it crosses zero.

\begin{figure}[!tb]
  \centering
  \includegraphics[width=0.49\textwidth]{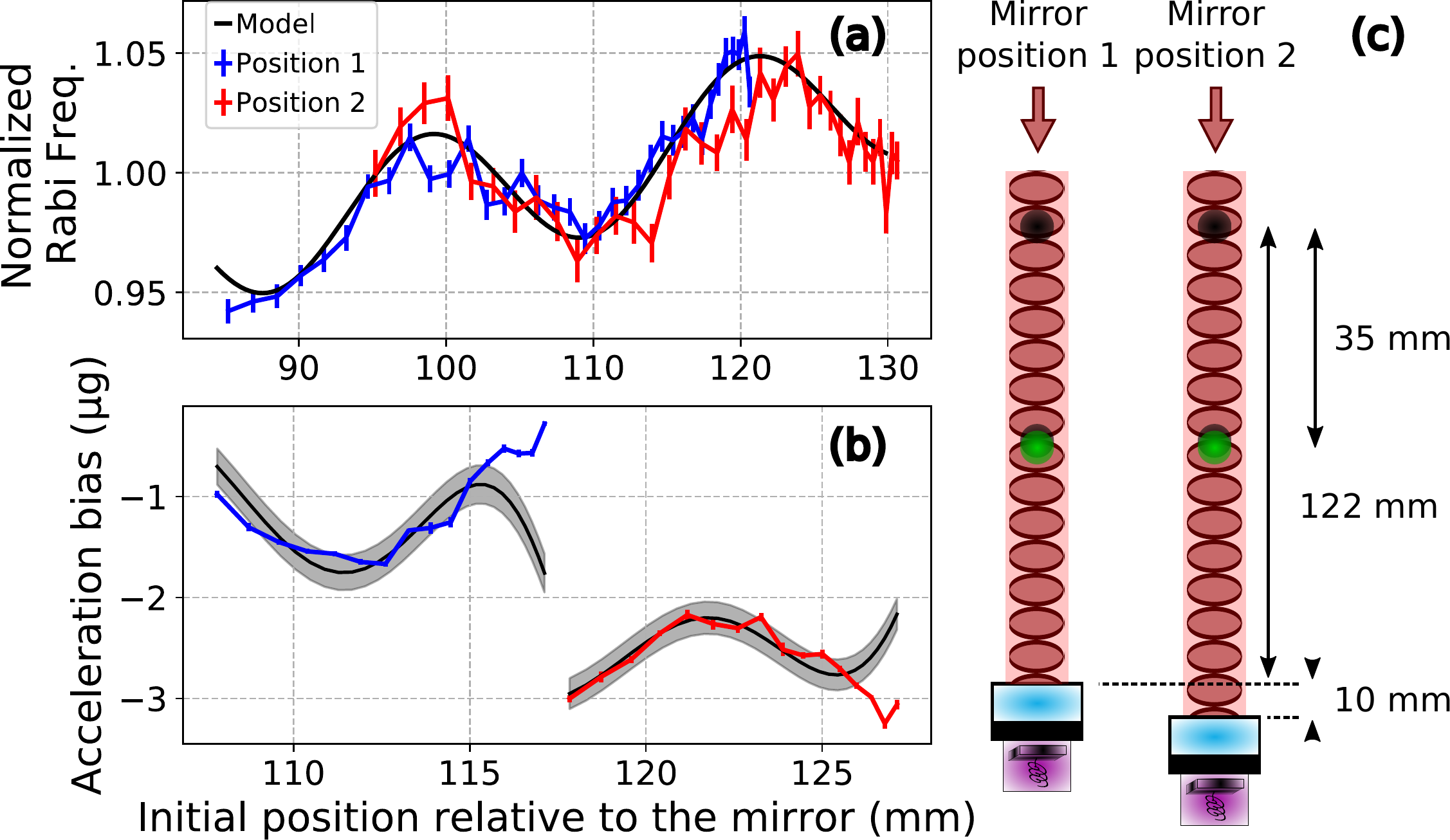}
  \caption{(a) Measurements of the normalized Rabi frequency ($\chi_{\rm eff}/\chi_{0,0}$) as a function of the cloud position. The blue (red) points correspond to mirror position 1 (2). The solid black curve corresponds to our model for the IQ modulator. (b) Measurements of the induced frequency-step bias due to the spatial variation of the Rabi frequency for an interrogation time $T = 20$ ms. (c) Schematic of the setup. The atom-mirror distance is varied by letting the cloud (black dot) fall for different heights over a maximum of 35 mm (constrained by our vacuum system). The mirror was then moved from position 1 to position 2 between data sets.}
  \label{fig:BiasVsMirrorPosition}
\end{figure}

To confirm the validity of our model for the IQ modulator and its affects on the AI, we measured the spatial variation of the Rabi frequency. Figure \ref{fig:BiasVsMirrorPosition}(a) shows measurements of this variation as a function of the cloud position relative to the mirror. The central cloud position was varied by changing the time-of-flight before the AI. We then measured the Rabi frequency at a fixed position by varying the duration of the pulse and fitting the resulting Rabi oscillations. The normalized Rabi frequency ($\chi_{\rm eff}/\chi_{0,0}$) is estimated by dividing the measurements by their mean value. Our model for the Rabi frequency (see \App \ref{sec:ResidualLineEffects}) shows excellent agreement with the data, and demonstrates that the residual laser lines spatially modulate the Raman coupling by $\sim 5\%$ peak-to-peak. This corresponds to a factor of $\sim 3.5$ reduction compared to a phase modulator. We expect this have a small effect on the AI fringe contrast, but we did not observe significant contrast modulations within our operating parameters.

Based on our model, we estimate a phase shift of $\Delta\Phi \simeq 6$ mrad for typical experimental conditions (see \Fig \ref{fig:IQvsPM}), which corresponds to an acceleration bias of 95 n$g$. Compared to a phase modulator, this systematic is reduced by a factor of $\sim 100$. However, it is challenging to measure phase shifts at this level (for our conditions, it would require averaging for several hours). To enhance the phase shift due to residual laser lines, we modified the Doppler-compensation protocol for the AI. Instead of using a frequency chirp $\alpha \simeq k_{\rm eff} g$, we applied a sequence of phase-continuous frequency steps of $\alpha (T + 2\tau)$ between each Raman pulse (similar to the first atomic gravimeters \cite{Peters2001}). With this method, the Doppler shift is cancelled between pulses, but remains during the pulses. This creates a slight imbalance between the phase contributions from atomic motion and the laser frequency (see \App \ref{sec:FrequencyProfile}). The result is a phase shift $\Delta\Phi_{\rm FS}$ that is proportional to the difference between the Rabi frequencies $\chi_{\rm eff}^{(1)}$ and $\chi_{\rm eff}^{(3)}$ at each $\pi/2$-pulse:
\begin{subequations}
\label{ScaleFactorPhase}
\begin{align}
  \Delta\Phi_{\rm FS} & = \alpha (T + 2\tau)\left[ \frac{\tan\Big( \frac{\chi_{\rm eff}^{(3)}\tau}{2} \Big)}{\chi_{\rm eff}^{(3)}} - \frac{\tan\Big(\frac{\chi_{\rm eff}^{(1)}\tau}{2}\Big)}{\chi_{\rm eff}^{(1)}} \right], \\
  & \simeq \alpha (T + 2\tau)(\pi/2-1) \big(\chi_{\rm eff}^{(3)} - \chi_{\rm eff}^{(1)}\big) / \chi_{0,0}^2.
\end{align}
\end{subequations}
Here, the last line is a first-order expansion about $\chi_{\rm eff}^{(1)} = \chi_{\rm eff}^{(3)} = \chi_{0,0}$. This phase shift is much larger than $\Delta\Phi$ (typically several hundred mrad) due to the spatial variation of the Rabi frequency. Hence, this technique allows us to validate the influence of the IQ modulator on the AI.

Figure \ref{fig:BiasVsMirrorPosition}(b) shows measurements of the acceleration bias due to the laser frequency steps ($\Delta\Phi_{\rm FS}/k_{\rm eff} T^2$) as a function the atom-mirror distance at the time of the first pulse. The data are first corrected for the two-photon light shift \cite{Gauguet2008} and subsequently compared with the predicted frequency-step bias using our spatially-varying Rabi frequency as input. We note that the atomic trajectories differ for interferometers employing $\pm k_{\rm eff}$ and this must be accounted for in the model. We obtain good agreement between the model and the data in both mirror positions when constant biases of $-0.9$ $\mu$g and $-1.4$ $\mu$g are added to the model for respectively mirror positions 1 and 2. Those biases are attributed to slight tilts of the collimator ($\sim 0.67$ mrad) and the mirror ($\sim 0.35$ mrad) relative to the vertical between the two positions. We also shifted the model by $-3.5$ mm on the position axis. This shift is attributed to effects related to the spatial and velocity distribution of the cloud and requires further study. We emphasize that the modulation amplitude and spatial period of the frequency-step bias show excellent agreement with predictions. These results verify our model for the Rabi frequency and, by extension, the phase shift $\Delta\Phi$ due to residual laser lines.

\section{Conclusion}
\label{sec:Conclusion}

We presented a dual-frequency modulated laser source with reduced parasitic sidebands, where the control of two independent optical signals is realized using microwave signals. The key component is a IQ modulator that enables optimal suppression of the carrier and the production of two single sidebands, with a rejection of the parasitic lines better than 20 dB. With this laser source, we demonstrated a complete cold-atom measurement sequence, including laser cooling, state preparation and atom interferometry, and detection, using only rf control. The laser system is tunable over a range of 15 GHz in typically 1 $\mu$s using a microwave waveform generator with large time-bandwidth \cite{Zhou2016}. This architecture can be employed for many cold atoms techniques like atom launching, high Doppler-shift compensations and zero-velocity atom interferometry \cite{Perrin2019}. The dual single-sideband modulation can be generalized to any multiple single-sideband generation, while limiting the influence of unwanted frequencies.

Our theoretical model predicts a systematic error due to residual laser lines below 100 n$g$ at $T = 20$ ms, which is confirmed by measurements of the Rabi frequency at different cloud positions. We also measured the influence of residual lines on an atomic gravimeter via the acceleration bias induced by a sequence of phase-continuous laser frequency steps. We anticipate that the contribution due to residual lines can be reduced to 10 n$g$ by further optimizing the operating parameters of the laser system (see \App \ref{sec:ResidualLineEffects}). The tunability of the IQ modulator implies that our architecture can also be adapted for other atomic species, such as potassium \cite{Menoret2011} or cesium \cite{Antoni-Micollier2018}, or upgraded to operate with a single diode. The simplicity of our architecture, along with the dramatic improvement in accuracy over phase-modulator-based designs, makes it a suitable alternative for systems employing two phase-locked lasers \cite{Oh2016, Li2017}. It is also easily transportable---making it ideal for a large range of onboard applications, including mobile gravity surveys \cite{Bidel2020}, inertial navigation \cite{Cheiney2018}, or fundamental physics in Space \cite{Aguilera2014}.

\begin{acknowledgments}

This work is supported by the French national agencies ANR (l'Agence Nationale pour la Recherche) and DGA (D\'{e}l\'{e}gation G\'{e}n\'{e}rale de l'Armement) under grant no.~ANR-17-ASTR-0025-01, and ESA (European Space Agency) under grant NAVISP no.~4000126014/18/NL/MP. We thank B. Gouraud for fruitful discussions, and the team at iXblue Photonics who developed the IQ modulator and assisted with its integration.

\end{acknowledgments}

\appendix

\section{Model of the electric field produced by CS-DSSB Modulation}
\label{sec:IQModel}

In this appendix, we present a derivation of the modulated optical field obtained using the Carrier Suppressed-Double Single Sideband (CS-DSSB) modulation technique in an electro-optic IQ modulator. For the purposes of this derivation, we use a plane wave at the input of the IQ modulator: $E_0 e^{i(\omega_0 t + \phi_0)}$, with a carrier frequency $\omega_0$, optical phase $\phi_0$, and electric field amplitude $E_0$. We ignore the spatial part of the field for the moment, but we discuss its effects in \App \ref{sec:ResidualLineEffects}. Following the scheme shown in \Fig \ref{fig:Architecture}(b), injecting two rf signals in phase-quadrature into the IQ modulator, the electric field at the output of the two sub-MZIs can be written as
\begin{widetext}
\begin{subequations}
\begin{align}
  E_1(t) & = \frac{E_0}{4} e^{i(\omega_0 t + \phi_0)} \left( e^{i[\beta_1 \cos(\Omega_1 t + \phi_1) + \beta_2 \cos(\Omega_2 t + \phi_2) + \Delta\Phi_1/2]} + e^{-i[\beta_1 \cos(\Omega_1 t + \phi_1) + \beta_2 \cos(\Omega_2 t + \phi_2) + \Delta\Phi_2/2]} \right), \\
  E_2(t) & = \frac{E_0}{4} e^{i(\omega_0 t + \phi_0)} \left( e^{i[\beta_1 \sin(\Omega_1 t + \phi_1) + \beta_2 \sin(\Omega_2 t + \phi_2) + \Delta\Phi_1/2]} + e^{-i[\beta_1 \sin(\Omega_1 t + \phi_1) + \beta_2 \sin(\Omega_2 t + \phi_2) + \Delta\Phi_2/2]} \right).
\end{align}
\end{subequations}
Here, $\Omega_1$ and $\Omega_2$ represent the two rf frequencies, $\phi_1$ and $\phi_2$ are arbitrary phases, and $\beta_1$ and $\beta_2$ are the respective rf modulation depths given by $\beta_i = \pi(V_i/V_\pi)$, where $V_i$ is the signal amplitude and $V_\pi$ is the half-wave voltage of the two sub-MZIs. This quantity depends on the dimensions of the optical waveguide, which is optimized to reduce $V_\pi$ as much as possible. $\Delta\Phi_1$ and $\Delta\Phi_2$ are the phase differences between the arms of each sub-MZI, and the main MZI contains a phase difference $\Delta\Phi_3$. All of these phases are controlled by DC bias voltages within each MZI. The total electric field at the output of the IQ modulator is $E_{\rm IQ} = E_1 e^{i\Delta\Phi_3/2} + E_2 e^{-i\Delta\Phi_3/2}$. In CS-DSSB mode, we tune the phases $\Delta\Phi_1 = \Delta\Phi_2 = \pi$ in order to suppress the carrier. Using the Jacobi-Anger expansion, it follows that
\begin{subequations}
\label{E12Expanded}
\begin{align}
  E_1(t) & = \frac{E_0}{2} e^{i(\omega_0 t + \phi_0)}
  \sum_{n,m \in \mathbb{Z}} \cos\left((n+m+1)\frac{\pi}{2}\right) J_n(\beta_1) J_m(\beta_2) e^{i [(n \Omega_1 + m \Omega_2) t + n\phi_1 + m\phi_2]}, \\
  E_2(t) & = \frac{E_0}{2} e^{i(\omega_0 t + \phi_0)}
   \sum_{n,m \in \mathbb{Z}} \cos\left((n+m+1)\frac{\pi}{2}\right) J_n(\beta_1) J_m(\beta_2) e^{i [(n \Omega_1 + m \Omega_2) t + n\phi_1 + m\phi_2]} e^{-i(n+m)\pi/2},
\end{align}
\end{subequations}
where $J_n(z)$ is a Bessel function of the first kind. One can easily verify that the amplitude of the carrier frequency (corresponding to indices $n = m = 0$) vanishes for both $E_1$ and $E_2$. Combining \Eqs \eqref{E12Expanded} in quadrature (\ie with $\Delta\Phi_3 = -\pi/2$) we can suppress the lower sidebands at the output of the IQ modulator. It is convenient to write this electric field in the following form
\be
  \label{EIQ}
  E_{\rm IQ}(t) = E_0 e^{i(\omega_0 t + \phi_0)} \sum_{n,m \in \mathbb{Z}} A_{n,m}^{\rm IQ} e^{i \big(\Omega_{n,m}^{\rm IQ} t + \phi_{n,m}^{\rm IQ}\big)},
\ee
where the amplitude, frequency, and relative phase of each line are
\begin{subequations}
\begin{align}
  \label{AIQ}
  A_{n,m}^{\rm IQ} & = \cos\left((n+m+1)\frac{\pi}{2}\right) \cos\left((n+m-1)\frac{\pi}{4}\right) J_n(\beta_{1}) J_m(\beta_{2}), \\
  \label{OmegaIQ}
  \Omega_{n,m}^{\rm IQ} & = n\Omega_1 + m\Omega_2, \\
  \label{phiIQ}
  \phi_{n,m}^{\rm IQ} & = n\phi_1 + m\phi_2 - (n+m)\frac{\pi}{4}.
\end{align}
\end{subequations}
According to \Eq \eqref{AIQ}, all harmonics corresponding to even $n+m$ or odd $(n+m-1)/2$ are suppressed. For example, the harmonics with indices $(n,m) = (-2,0),~(-1,0),~(0,0),~(0, -1),~(1,1),~(1,2)$ are all suppressed due to our choice of $\Delta\Phi_3 = -\pi/2$. The principal lines used for laser cooling and interferometry in our case correspond to $(n,m) = (0,1)$ and $(1,0)$. The electric field at the output of the PPLN is proportional to the square of the input field: $E_{\rm PPLN}(t) = \eta\epsilon_0\chi^2 E_{\rm IQ}^2(t)$, where $\eta$ is the efficiency of second-harmonic generation (SHG), $\epsilon_0$ is the vacuum permittivity, and $\chi$ is the susceptibility of the medium. Squaring \Eq \eqref{EIQ}, we obtain:
\begin{align}
  \label{EPPLN1}
  E_{\rm PPLN}(t) & = \eta\epsilon_0 \chi^2 E_0^2 e^{i(2\omega_0 t + 2\phi_0)} \sum_{n,m,l,k \in \mathbb{Z}} \cos\left((n+m+1)\frac{\pi}{2}\right) \cos\left((n+m-1)\frac{\pi}{4}\right) \cos\left((l+k+1)\frac{\pi}{2}\right) \\
  & \times \cos\left((l+k-1)\frac{\pi}{4}\right)
  J_n(\beta_1) J_m(\beta_2) J_l(\beta_1) J_k(\beta_2) e^{i \left[\big((n+l) \Omega_1 + (m+k) \Omega_2\big) t + (n+l) \phi_1 + (m+k) \phi_2 - (n+m+l+k)\pi/4 \right]}. \nonumber
\end{align}
To simplify this expression, we define new indices $N = n+l$ and $M=m+k$, and substitute them in \Eq \eqref{EPPLN1}. The electric field after SHG can then be written as
\be
  \label{EPPLN2}
  E_{\rm PPLN}(t) = \eta\epsilon_0 \chi^2 E_0^2 e^{i(2\omega_0 t + 2\phi_0)} \sum_{N,M \in \mathbb{Z}} A_{N,M}^{\rm PPLN} e^{i \left(\Omega_{N,M}^{\rm PPLN} t + \phi_{N,M}^{\rm PPLN}\right)},
\ee
where the amplitude, frequency, and relative phase are
\begin{subequations}
\begin{align}
  \label{APPLN}
  A_{N,M}^{\rm PPLN} & = \sum_{n,m\in\mathbb{Z}} \cos\left((n+m+1)\frac{\pi}{2}\right) \cos\left((N+M-n-m+1)\frac{\pi}{2}\right) \cos\left((n+m-1)\frac{\pi}{4}\right) \\
  & \times \cos\left((N+M-n-m-1)\frac{\pi}{4}\right) J_n(\beta_1) J_{N-n}(\beta_1) J_m(\beta_2) J_{M-m}(\beta_2), \nonumber \\
  \label{OmegaPPLN}
  \Omega_{N,M}^{\rm PPLN} & = N\Omega_1 + M\Omega_2, \\
  \label{phiPPLN}
  \phi_{N,M}^{\rm PPLN} & = N\phi_1 + M\phi_2 - (N+M)\frac{\pi}{4}.
\end{align}
\end{subequations}
\end{widetext}
Equation \eqref{APPLN} determines the relative power ratio between the harmonics and allows us to cancel constant light shifts by precisely controlling the intensity ratio of the two principal lines \cite{LeGouet2008}. We emphasize that the sum over the individual line indices $n,m$ appears only in the amplitude, while the frequency and phase of each second-harmonic is determined by indices $N,M$ from the SHG process. The principal Raman lines in our case correspond to $(N,M) = (2,0)$ and $(1,1)$. Although all other lines are suppressed, each pair of lines separated by $\delta_{\rm HF}$ is resonant with a two-photon Raman transition. Using \Eq \eqref{OmegaPPLN} and the fact that $\Omega_2 = \Omega_1 + \delta_{\rm HF}$, this condition can be summarized as
\begin{align}
  \delta_{\rm HF}
  & = \Omega_{N',M'}^{\rm PPLN} - \Omega_{N,M}^{\rm PPLN}, \\
  & = (N'-N)\Omega_1 + (M'-M)(\Omega_1 + \delta_{\rm HF}). \nonumber
\end{align}
It follows that only pairs $(\Omega_{N',M'}^{\rm PPLN}, \Omega_{N,M}^{\rm PPLN})$ with $N' = N-1$ and $M' = M+1$ can drive Raman transitions. The phase difference between any of these Raman lines can be deduced from \Eq \eqref{phiPPLN}
\be
  \label{phiRaman}
  \Delta\phi = \phi_{N',M'}^{\rm PPLN} - \phi_{N,M}^{\rm PPLN} = \phi_2 - \phi_1,
\ee
which is equal to the phase difference between the two input rf signals. This key property enables precise control of the atom interferometer phase via the optical Raman interaction.

\section{Effects of residual laser lines on the atom interferometer}
\label{sec:ResidualLineEffects}

Each pair of residual laser lines separated by $\delta_{\rm HF}$ is resonant with a two-photon Raman transition and can lead to a parasitic phase shift in the atom interferometer, as illustrated in \Fig \ref{fig:ParasiticAI}. Using our model for the electric field in the previous section, here we develop a model for the phase shift and contrast loss in the atom interferometer produced by the IQ modulator. We adopt the approach of \Ref \citenum{Carraz2012}, where the impact of additional laser lines was previously studied in the case of an electro-optic phase modulator.

The counter-propagating Rabi frequency associated with any two pairs of lines is proportional to the product of their electric field amplitudes, and inversely proportional to the detuning from the excited state. For all resonant pairs of lines $(\Omega_{2+p,q}^{\rm PPLN}, \Omega_{1+p,1+q}^{\rm PPLN})$, where the indices $(p,q)$ denote the offset from the principal Raman lines $(\Omega_{2,0}^{\rm PPLN}, \Omega_{1,1}^{\rm PPLN})$, the Raman coupling parameter can be written as
\be
  \Lambda_{p,q} \sim \frac{E_{2+p,q}(t) E_{1+p,1+q}^*(t - 2z/c)}{\Delta_{\rm R} + p\Omega_1 + q\Omega_2},
\ee
where $E_{N,M}(t) = E_0 e^{i 2\omega_0 t} A_{N,M}^{\rm PPLN} e^{i (\Omega_{N,M}^{\rm PPLN} t + \phi_{N,M}^{\rm PPLN})}$ is an electric field amplitude, $z$ is the distance between the atoms and the mirror at time $t$, and $2z/c$ is the round-trip time required for the light to reflect off the mirror. It follows that
\begin{subequations}
\begin{align}
  \Lambda_{p,q} & \sim e^{i(k_{\rm eff} z - \delta_{\rm HF} t - \Delta\phi)} \chi_{p,q} e^{i(p\Delta k_1 + q\Delta k_2) z}, \\
  \chi_{p,q} & \propto \frac{A_{2+p,q}^{\rm PPLN} A_{1+p,1+q}^{\rm PPLN}}{\Delta_{\rm R} + p\Omega_1 + q\Omega_2},
\end{align}
\end{subequations}
where $\chi_{p,q}$ is a Rabi frequency and the effective wavevector is
\begin{align}
  k_{\rm eff}
  & = (2\omega_0 + \Omega_{2,0}^{\rm PPLN} + 2\omega_0 + \Omega_{1,1}^{\rm PPLN})/c \nonumber \\
  & = (4\omega_0 + 3\Omega_1 + \Omega_2)/c.
\end{align}
The first term in $\Lambda_{p,q}$ describes the energy ($\hbar \delta_{\rm HF}$), momentum ($\hbar k_{\rm eff}$), and phase ($\Delta\phi$) transferred to the atoms by the principal Raman lines during each pulse. These lines are associated with the principal Rabi frequency $\chi_{0,0} \equiv \pi/2\tau$, where $\tau$ is the $\pi/2$-pulse duration. For other pairs of lines, the energy and phase are identical, but due to additional spatial harmonics present in the field, the momentum transfer is modified to $\hbar(k_{\rm eff} + p\Delta k_1 + q\Delta k_2)$, where $\Delta k_1 = 2\Omega_1/c$ and $\Delta k_2 = 2\Omega_2/c$. This slightly different momentum kick for each pair of laser lines is the origin of the parasitic phase shift in the atom interferometer.

The Rabi frequencies $\chi_{p,q}$ associated with each resonant pair of laser lines can be determined from experimental parameters using the following ratio
\begin{align}
  \label{chinm}
  \frac{\chi_{p,q}}{\chi_{0,0}} & = \frac{A_{2+p,q}^{\rm PPLN} A_{1+p,1+q}^{\rm PPLN}}{A_{2,0}^{\rm PPLN} A_{1,1}^{\rm PPLN}} \\
  & \times \left( \frac{\frac{1}{\Delta_{\rm R2} + p\Omega_1 + q\Omega_2} + \frac{1}{3(\Delta_{\rm R1} + p\Omega_1 + q\Omega_2)}}{\frac{1}{\Delta_{\rm R2}} + \frac{1}{3\Delta_{\rm R1}}} \right), \nonumber
\end{align}
where $\Delta_{\rm R2} = \Delta_{\rm R}$ and $\Delta_{\rm R1} = \Delta_{\rm R} + 2\pi \times 157$ MHz are the single-photon Raman detunings from the $\ket{F=2} \to \ket{F'=2}$ and $\ket{F'=1}$ transitions in $^{87}$Rb, respectively. Table \ref{tab:RabiFreqs} provides a list of the Rabi frequencies and detunings used in our calculations.

\begin{table}[!tb]
  \begin{ruledtabular}
  \begin{tabular}{rrrrrrr}
        $p$ & $q$ & $|\chi_{p,q}/\chi_{0,0}|$ & sign($\chi_{p,q}$) & $\Omega(p,q)$ & $\Delta_{\rm R1}(p,q)$ & $\Delta_{\rm R2}(p,q)$ \\
            &     &           (dB) &                    &         (GHz) & (GHz) & (GHz) \\
    \hline
       -2 &  0 & -25.5 &  1 & -1.91 & -2.63 & -2.79 \\
       -1 &  0 & -23.2 & -1 & -0.96 & -1.67 & -1.83 \\
       -1 &  1 & -18.5 & -1 &  6.84 &  6.12 &  5.96 \\
        0 &  0 &   0.0 &  1 &  0.00 & -0.72 & -0.88 \\
        1 & -1 & -18.5 & -1 & -6.84 & -7.56 & -7.71 \\
        1 &  0 & -15.6 &  1 &  0.96 &  0.24 &  0.08 \\
        2 &  0 & -27.4 &  1 &  1.91 &  1.19 &  1.03 \\
  \end{tabular}
  \end{ruledtabular}
  \caption{Table of Rabi frequencies and detunings for $\Omega_1/2\pi = 0.96$ GHz, $\Omega_2/2\pi = 7.79$ GHz, $\Delta_{\rm R}/2\pi = -0.88$ GHz, and line intensities measured from \Fig \ref{fig:IQSpectra}(b). Here $\Omega(p,q) = p\Omega_1 + q\Omega_2$, and $\Delta_{\rm R1,2}(p,q) = \Delta_{\rm R1,2} + p\Omega_1 + q\Omega_2$. Other quantities are as defined in the text.}
  \label{tab:RabiFreqs}
\end{table}

During each Raman pulse, the light-matter interaction imprints a phase on the atoms that depends on the position $z$ relative to the retro-reflection mirror. The effective Rabi frequency $\chi_{\rm eff}$ determines the strength of this interaction. In the presence of additional laser lines, this quantity is given by the sum over all Rabi frequencies---each coupled with a phase term describing the modified momentum transfer
\be
  \chi_{\rm eff}(z) = \sum_{p,q \in \mathbb{Z}} \chi_{p,q} e^{i(p\Delta k_1 + q\Delta k_2)z}.
\ee
This sum results in a spatially-varying Rabi frequency due to the interference between different spatial harmonics. In addition to the rf phase $\Delta\phi$, the atoms are imprinted with the following phase during each Raman pulse
\be
  \varphi(z) = {\rm arg}\left( \chi_{\rm eff}(z) \right).
\ee
In a three-pulse Mach-Zehnder interferometer, the total phase shift is
\be
  \Delta\Phi = \varphi(z_{\rm A}) - \varphi(z_{\rm B}) - \varphi(z_{\rm C}) + \varphi(z_{\rm D}),
\ee
where $z_{\rm A}, \ldots, z_{\rm D}$ denote the vertices of the interferometer at times $t = {\rm TOF}$, ${\rm TOF} + T$, and ${\rm TOF} + 2T$ (see \Fig \ref{fig:ParasiticAI}). In a gravimeter configuration, where the atoms fall within a vertically-aligned laser beam at a constant acceleration $a_z = -g$, these vertices are
\begin{subequations}
\label{AIVertices}
\begin{align}
  z_{\rm A} & = z_{\rm M} + v_0 {\rm TOF} + \frac{1}{2} a_z ({\rm TOF})^2, \\
  z_{\rm B} & = z_{\rm A} + (v_0 + a_z {\rm TOF}) T + \frac{1}{2} a_z T^2, \\
  z_{\rm C} & = z_{\rm A} + (v_0 + a_z {\rm TOF} + v_{\rm rec}) T + \frac{1}{2} a_z T^2, \\
  z_{\rm D} & = z_{\rm A} + \left( v_0 + a_z {\rm TOF} + \frac{1}{2} v_{\rm rec} \right) (2T) + \frac{1}{2} a_z (2T)^2.
\end{align}
\end{subequations}
Here, $v_0$ is the initial velocity of the atomic cloud, TOF is the time-of-flight before the first Raman pulse, $z_{\rm M}$ is the position of the mirror (nominally $122$ mm below the initial cloud position), $T$ is the interrogation time, and $v_{\rm rec} = \hbar k_{\rm eff}/M$ is the two-photon recoil velocity.

\begin{figure}[!tb]
  \centering
  \includegraphics[width=0.47\textwidth]{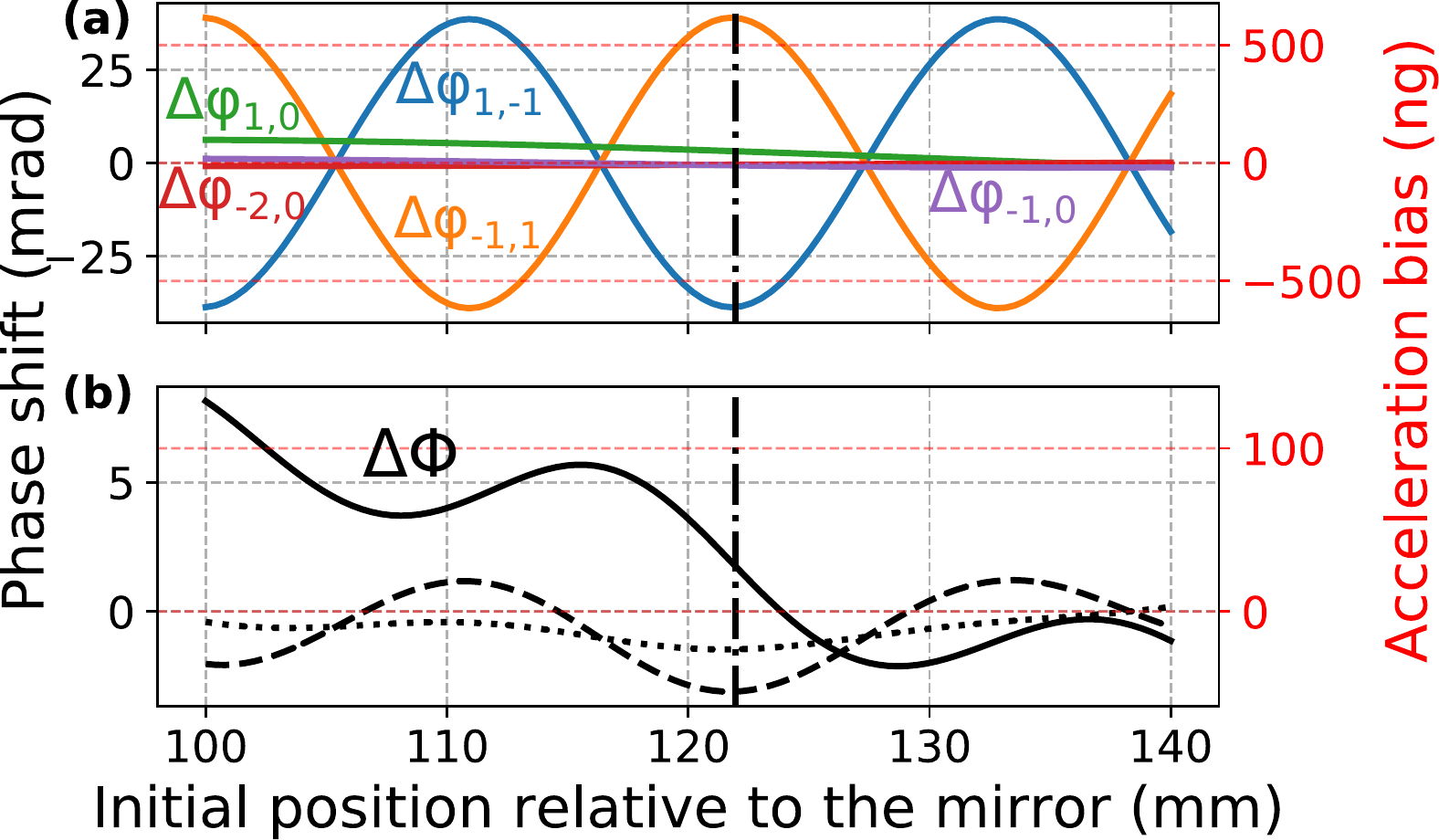}
  \caption{(a) Contribution of each parasitic Raman transition to the phase shift as a function of the atom-mirror distance $z_{\rm M}$. (b) Predicted AI phase shift due to residual parasitic lines for the current laser parameters (solid black curve), with optimized $\Omega_1$ (dashed curve), and with optimized modulation depths $\beta_1$ and $\beta_2$ (dotted line). AI parameters: TOF = 15 ms, $T = 20$ ms, $\Delta_{\rm R}/2\pi = -0.88$ GHz.}
  \label{fig:IQPhaseShift}
\end{figure}

Figure \ref{fig:IQPhaseShift} shows the predicted phase shift $\Delta\Phi$ due to residual lines as a function of the atom-mirror distance $z_{\rm M}$. For each line, we computed the phase contribution $\Delta\Phi_{p,q}$ using the corresponding parameters listed in Table \ref{tab:RabiFreqs}. It is clear from \Fig \ref{fig:IQPhaseShift}(a) that the largest contributors ($\Delta\Phi_{-1,1}$ and $\Delta\Phi_{1,-1}$) have opposite sign---hence their sum tends to zero. We emphasize that this is a direct result of our frequency-doubled CS-DSSB architecture \cite{Templier2021}. The total phase shift depicted in \Fig \ref{fig:IQPhaseShift}(b) presents locally an amplitude of $\sim 10$ mrad (162 n$g$) and for a constant offset of $\sim 2.3$ mrad (10 n$g$). The main contribution to this shift derives from the Rabi frequency $\chi_{1,0}$, where the Raman detunings are relatively small ($\Delta_{\rm R1}(1,0) \simeq 240$ MHz and $\Delta_{\rm R2}(1,0) \simeq 80$ MHz). By increasing $\Omega_1$ to 1.1 GHz, the acceleration shift is decreased and can be locally approximated as a sinusoidal oscillation with an amplitude and a constant offset of 68 and $-9$ n$g$, respectively. Further optimization of the modulation depths to $\beta_1 = 0.573$ and $\beta_2 = 0.255$ reduces the contributions from $\Delta\Phi_{-1,1}$ and $\Delta\Phi_{1,-1}$---resulting in an amplitude and offset of 26 and $-11$ n$g$, respectively.

\section{Measurement of atom-mirror distance using optical Raman spectroscopy}
\label{sec:MirrorDistance}

\begin{figure}[!tb]
  \centering
  \includegraphics[width=0.49\textwidth]{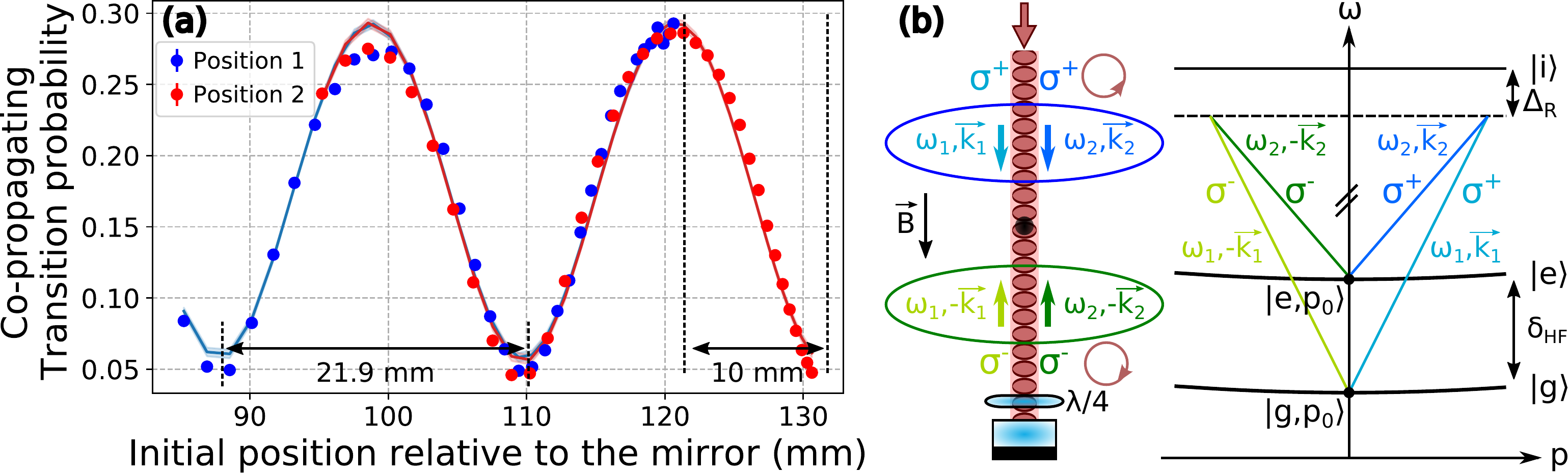}
  \caption{(a) Measurements of the co-propagating transition probability between $\ket{F,m_F=0}$ states as a function of the atom-mirror distance. Red (blue) points correspond to the nominal (shifted) mirror position. The solid curves are sinusoidal fits to the data. (b) Schematic of the experiment and level diagram for co-propagating transitions.}
  \label{fig:ProbVsMirrorPosition}
\end{figure}

To precisely determine the initial position of the atomic cloud relative to the reference mirror, as well as the residual launch velocity after the preparation stage, we took advantage of an interference effect using optical Raman spectroscopy. Two counter-propagating pairs of co-propagating Raman beams excite velocity-insensitive transitions, as shown in \Fig \ref{fig:ProbVsMirrorPosition}(b). These two orthogonally-polarized pairs of beams are simultaneously resonant, but the retro-reflected pair accumulates an additional phase proportional to the round-trip distance to the mirror. This results in a spatial modulation of the Rabi frequency analogous to the previous analysis with additional laser lines. Adding the electric fields for each pair of beams (ignoring residual lines), the effective Rabi frequency can be shown to be
\begin{align}
  \chi_{\rm co}(z)
  & = \chi_{\rm i} e^{i\Delta k z} + \chi_{\rm r} e^{-i\Delta k (z - 2 z_{\rm M})}, \nonumber \\
  & = 2 |\chi| e^{i\Delta k z_{\rm M}} \sin \Delta k (z - z_{\rm M}),
\end{align}
where $\chi \equiv \chi_{\rm i} = -\chi_{\rm r}$ are Rabi frequencies for the incident and reflected beams, and $\Delta k = \delta_{\rm HF}/c \simeq 2\pi \times 22.8$ m$^{-1}$ is the effective wavenumber for co-propagating transitions. This provides a powerful method to measure the initial atom-mirror distance $z_{\rm M}$, as well as the initial velocity of the cloud $v_0$.

Figure \ref{fig:ProbVsMirrorPosition}(a) shows measurements of the co-propagating transitions probability between magnetically-insensitive states as a function of the atom-mirror distance. We used a similar measurement strategy as for \Fig \ref{fig:BiasVsMirrorPosition}: the atom-mirror distance is varied using the time-of-flight before the Raman pulse, and we repeated the measurement for two different mirror positions. Here, we fixed the intensity and duration of the $\pi$-pulse, and recorded Raman spectra for each time-of-flight. We extracted the peak amplitude from fits to these spectra \cite{Templier2021}, which provide a direct measurement of the co-propagating transition probability shown in \Fig \ref{fig:ProbVsMirrorPosition}(a). This probability varies spatially as $P_{\rm co}(z) \propto \sin^2[\chi_{\rm co}(z) \tau]$, which features a spatial period of $\pi/\Delta k = 21.9$ mm. Assuming a parabolic trajectory for the cloud $z(t) = v_0 t - \frac{1}{2} g t^2$, a fit to these data provides a precise estimate of the initial atom-mirror distance $z_{\rm M} = 121.96(18)$ mm, as well as the initial vertical velocity of $v_0 = -15.5(3.1)$ mm/s.

\begin{figure}[!b]
  \centering
  \includegraphics[width=0.45\textwidth]{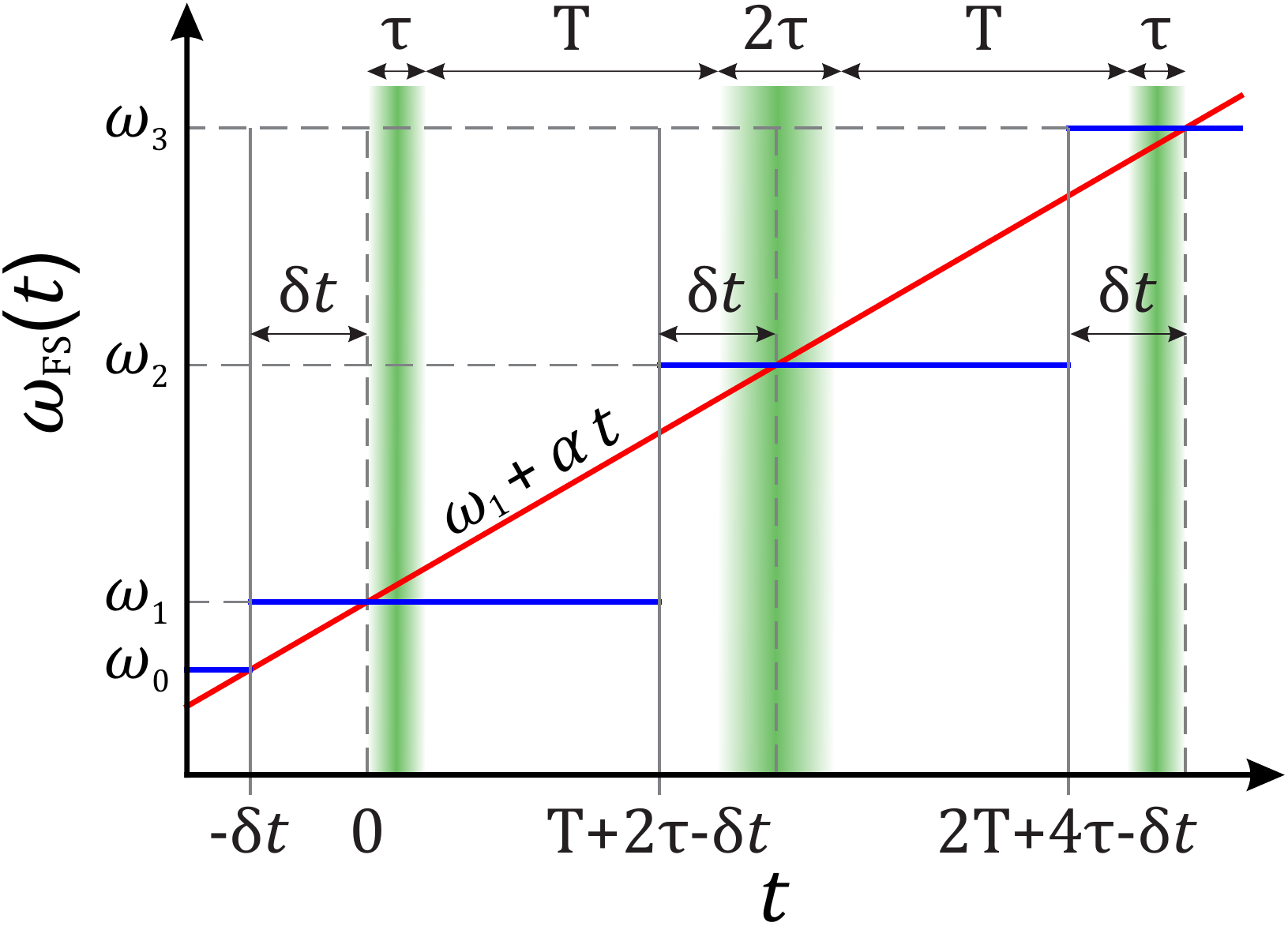}
  \caption{Timing diagram for frequency steps used in experiments (blue), and the corresponding frequency chirp (red) usually employed in atomic gravimeters.}
  \label{fig:FrequencyProfile}
\end{figure}

\begin{widetext}
\section{The AI phase shift due to phase-continuous frequency steps}
\label{sec:FrequencyProfile}

To amplify the effects of residual laser lines while also compensating the Doppler effect, we applied a series of phase-continuous frequency steps to the Raman frequency. This simulates a frequency chirp \emph{between} pulses, but maintains a constant frequency \emph{during} the pulses. Figure \ref{fig:FrequencyProfile} shows the frequency profile $\omega_{\rm FS}(t)$ used in the experiment, which is described by
\be
  \label{omegaFS}
  \omega_{\rm FS}(t) = \left\{ \begin{array}{ll}
    \omega_0, & t < -\delta t \\
    \omega_1, & -\delta t \le t < T + 2\tau - \delta t \\
    \omega_2, & T + 2\tau - \delta t \le t < 2T + 4\tau - \delta t \\
    \omega_3. & t \ge 2T + 4\tau - \delta t
  \end{array} \right.
\ee
Here, $\omega_0$ is an arbitrary frequency, $\omega_n$ is the frequency during the $n^{\rm th}$ Raman pulse, and $\delta t$ is a small temporal offset that defines when the frequency steps occur before each pulse. Since $\omega_{\rm FS}(t)$ is a phase-continuous frequency profile, in order to avoid undesired AI phase shifts proportional to $\delta t$ the frequency steps must be triggered symmetrically with respect to the center of the interferometer at $t = T + 2\tau$ (as shown in \Fig \ref{fig:FrequencyProfile}). The resulting phase shift due to the laser is given by
\be
  \label{philas1}
  \phi_{\rm las} = \int g_s(t) \omega_{\rm FS}(t) \dd t.
\ee
Here, $g_s(t)$ is the generalized sensitivity function \cite{Bonnin2015}:
\be
  \label{g(t)}
  g_s(t) = \left\{ \begin{array}{ll}
    0, & t \le 0 \mbox{ or } t > 2T+4\tau \\
    -\frac{\sin(\chi_{\rm eff}^{(1)} t)}{\sin(\chi_{\rm eff}^{(1)} \tau)}, & 0 < t \leq \tau \\
    -1, & \tau < t \leq T+\tau \\
    \frac{\sin(\chi_{\rm eff}^{(2)} (t-T-2\tau))}{\sin(\chi_{\rm eff}^{(2)} \tau)}, & T+\tau < t \leq T+3\tau \\
    +1, & T+3\tau < t \leq 2T+3\tau \\
    -\frac{\sin(\chi_{\rm eff}^{(3)} (t-2T-4\tau))}{\sin(\chi_{\rm eff}^{(3)} \tau)}, & 2T+3\tau < t \leq 2T+4\tau
  \end{array} \right.
\ee
where $\chi_{\rm eff}^{(n)}$ is the effective Rabi frequency during the $n^{\rm th}$ pulse. Evaluating \Eq \eqref{philas1} for the frequency profile in \Eq \eqref{omegaFS}, we obtain
\be
  \label{philas2}
  \phi_{\rm las} = (\omega_2 - \omega_1)T + (\omega_1 - 2\omega_2 + \omega_3)(\delta t - \tau) + \omega_3 \frac{\tan\big( \frac{\chi_{\rm eff}^{(3)} \tau}{2} \big)}{\chi_{\rm eff}^{(3)}} - \omega_1 \frac{\tan\big( \frac{\chi_{\rm eff}^{(1)} \tau}{2} \big)}{\chi_{\rm eff}^{(1)}}.
\ee
To compensate the Doppler shift $k_{\rm eff} g t$ during the AI, we step the frequency such that $\omega_n = \omega_{n-1} + \alpha (T+2\tau)$ with $\alpha = k_{\rm eff} g$. Under these conditions, the $2^{\rm nd}$ term in \Eq \eqref{philas2} vanishes and we obtain
\be
  \label{philas3}
  \phi_{\rm las} = \omega_1 \left[ \frac{\tan\big( \frac{\chi_{\rm eff}^{(3)} \tau}{2} \big)}{\chi_{\rm eff}^{(3)}} - \frac{\tan\big( \frac{\chi_{\rm eff}^{(1)} \tau}{2} \big)}{\chi_{\rm eff}^{(1)}} \right] + k_{\rm eff} g (T + 2\tau) \left[ T + 2\frac{\tan\big( \frac{\chi_{\rm eff}^{(3)} \tau}{2} \big)}{\chi_{\rm eff}^{(3)}} \right].
\ee

The kinematic component of the phase shift (\ie that due to atomic motion assuming a constant acceleration $-g$ and velocity $v_1$ at the first Raman pulse) is given by:
\begin{subequations}
\label{phikin}
\begin{align}
  \phi_{\rm kin}
  & = k_{\rm eff} \int g_s(t) (v_1 - g t) \dd t, \\
  & \simeq k_{\rm eff} v_1 \left[\frac{\tan\big( \frac{\chi_{\rm eff}^{(1)} \tau}{2} \big)}{\chi_{\rm eff}^{(1)}} - \frac{\tan\big( \frac{\chi_{\rm eff}^{(3)} \tau}{2} \big)}{\chi_{\rm eff}^{(3)}} \right] - k_{\rm eff} g (T+2\tau) \left[T + \frac{\tan\big( \frac{\chi_{\rm eff}^{(1)} \tau}{2} \big)}{\chi_{\rm eff}^{(1)}}  + \frac{\tan\big( \frac{\chi_{\rm eff}^{(3)} \tau}{2} \big)}{\chi_{\rm eff}^{(3)}} \right].
\end{align}
\end{subequations}
where the last line is a leading-order approximation. Summing \Eqs \eqref{philas3} and \eqref{phikin}, we obtain the total interferometer phase shift:
\be
  \label{phitot}
  \phi_{\rm tot} = (\omega_1 - k_{\rm eff} v_1) \left[ \frac{\tan\big( \frac{\chi_{\rm eff}^{(3)} \tau}{2} \big)}{\chi_{\rm eff}^{(3)}} - \frac{\tan\big( \frac{\chi_{\rm eff}^{(1)} \tau}{2} \big)}{\chi_{\rm eff}^{(1)}} \right]
  + k_{\rm eff} g (T + 2\tau) \left[ \frac{\tan\big( \frac{\chi_{\rm eff}^{(3)} \tau}{2} \big)}{\chi_{\rm eff}^{(3)}} - \frac{\tan\big( \frac{\chi_{\rm eff}^{(1)} \tau}{2} \big)}{\chi_{\rm eff}^{(1)}} \right].
\ee
In the ideal case of a frequency chirp that cancels the Doppler shift at all times during the interferometer, the laser and kinetic phases are equal and opposite when $\alpha = k_{\rm eff} g$---resulting in $\phi_{\rm tot} = 0$. In the case of phase-continuous frequency steps we obtain two terms. The first term in \Eq \eqref{phitot} vanishes when the Raman frequency during the first pulse is equal to the Doppler shift ($\omega_1 = k_{\rm eff} v_1$). The remaining term is equivalent to \Eq \eqref{ScaleFactorPhase} in the main text, which can be approximated as $k_{\rm eff} g T (\pi/2 - 1) (\chi_{\rm eff}^{(3)} - \chi_{\rm eff}^{(1)}) (2\tau/\pi)^2$. Since this quantity scales as the difference between Rabi frequencies, it is a powerful tool to amplify any imbalance between them. In our case, it is ideal to indirectly measure the effects of residual laser lines as they produce a spatial modulation in the Rabi frequency.

Equation \eqref{phitot} can also be interpreted as an imbalance between the AI scale factors for the laser and kinematic phases. The laser phase can be written as $\phi_{\rm las} = S_{\rm las} (\alpha/k_{\rm eff})$, where $S_{\rm las}$ is the laser scale factor. This phase aims to compensate the kinetic phase which, for a constant acceleration $a$, is $\phi_{\rm kin} = S_{\rm kin} a$. When employing phase-continuous frequency steps, $S_{\rm las}$ differs from the kinetic scale factor $S_{\rm kin}$:
\begin{subequations}
\label{ScaleFactors}
\begin{align}
  \frac{S_{\rm kin}}{k_{\rm eff}}
  & = (T + 2\tau) \left[T + \frac{\tan\big( \frac{\chi_{\rm eff}^{(1)} \tau}{2} \big)}{\chi_{\rm eff}^{(1)}}  + \frac{\tan\big( \frac{\chi_{\rm eff}^{(3)} \tau}{2} \big)}{\chi_{\rm eff}^{(3)}} \right], \\
  \frac{S_{\rm las}}{k_{\rm eff}}
  & = (T + 2\tau) \left[T + 2 \frac{\tan\big( \frac{\chi_{\rm eff}^{(3)} \tau}{2} \big)}{\chi_{\rm eff}^{(3)}} \right].
\end{align}
\end{subequations}
When operating at the Doppler-compensating chirp rate $\alpha = k_{\rm eff} a$, this leads to a bias on the acceleration measurement since the total phase shift $\phi_{\rm tot} = (S_{\rm kin} - S_{\rm las}) a \neq 0$.

\end{widetext}


\bibliography{References-Short}

\end{document}